\documentclass[11pt,a4paper]{article}
\usepackage[colorlinks=true, linkcolor=black!50!blue, urlcolor=blue, citecolor=blue, anchorcolor=blue]{hyperref}
\usepackage[font=small,labelfont=bf,margin=0mm,labelsep=period,tableposition=top]{caption}
\usepackage[a4paper,top=3cm,bottom=2.5cm,left=2.5cm,right=2.5cm,bindingoffset=0mm]{geometry}

\usepackage{graphicx,placeins}
\usepackage{float}
\usepackage{afterpage}
\usepackage{epsfig,cite}
\usepackage{amssymb}
\usepackage{amsmath}
\usepackage{dsfont}
\usepackage{multirow}
\usepackage{tabularx}
\usepackage{setspace}
\usepackage{url}
\usepackage{xcolor,colortbl}
\usepackage{float}
\usepackage{afterpage}
\usepackage{url}
\usepackage{hyperref}
\usepackage{booktabs}
\usepackage{mathrsfs}
\usepackage{subcaption}
\usepackage{bm}       

\usepackage{tikz}
\usepackage{tikz-3dplot}
\usepackage[compat=1.0.0]{tikz-feynman}

\usepackage{enumitem}
\usepackage{hyperref}
\usepackage{cite}

\usepackage{pifont}

\usetikzlibrary{shapes, arrows}
\usetikzlibrary{decorations.pathreplacing}
\usetikzlibrary{positioning, calc}
\tikzstyle{fitted} = [rectangle, minimum width=5cm, minimum height=1cm, text centered, draw=black, fill=red!30]
\tikzstyle{operations} = [rectangle, rounded corners, minimum width=2cm,text centered, draw=black, fill=red!30]
\tikzstyle{roundtext} = [rectangle, rounded corners, minimum width=2cm, minimum height=0.8cm, text centered, draw=black, fill=red!30]
\tikzstyle{n3py} = [rectangle, rounded corners, minimum width=3cm, minimum height=1cm, text centered, draw=black, fill=green!30]
\tikzstyle{myarrow} = [thick,->,>=stealth]
\tikzstyle{line} =[draw, -latex']
\tikzstyle{decision} = [diamond, draw, fill=red!20, text width=7.5em, text centered,  inner sep=0pt, minimum height=2em, aspect=4]
\tikzstyle{cloud} = [draw, ellipse,fill=green!20, minimum height=2em]
\tikzstyle{inout} = [rectangle, draw, fill=green!20, text width=9.5em, text centered, rounded corners, minimum height=2em, minimum width=10em]
\tikzstyle{block}=[rectangle, draw, fill=blue!20, text width=9.5em,
                   text centered, rounded corners, minimum height=2em,
                   minimum width=10em]

\definecolor{darkgreen}{rgb}{0.0, 0.5, 0.13}

\newcommand{\bea}{\begin{eqnarray}}
\newcommand{\eea}{\end{eqnarray}}
\newcommand{\bi}{\begin{itemize}}
\newcommand{\ei}{\end{itemize}}
\newcommand{\ben}{\begin{enumerate}}
\newcommand{\een}{\end{enumerate}}

\newcommand{\lc}{\left[}
\newcommand{\rc}{\right]}
\newcommand{\lp}{\left(}
\newcommand{\rp}{\right)}

\def\gsim{\mathrel{\rlap{\lower4pt\hbox{\hskip1pt$\sim$}}
    \raise1pt\hbox{$>$}}}         
\def\lsim{\mathrel{\rlap{\lower4pt\hbox{\hskip1pt$\sim$}}
    \raise1pt\hbox{$<$}}}         

\newcommand{\draft}[1]{}

\def\beq{\begin{equation}}
\def\eeq{\end{equation}}

\def\lapprox{\lower .7ex\hbox{$\;\stackrel{\textstyle <}{\sim}\;$}}
\def\gapprox{\lower .7ex\hbox{$\;\stackrel{\textstyle >}{\sim}\;$}}

\numberwithin{equation}{section}
\numberwithin{figure}{section}
\numberwithin{table}{section}

\usepackage{tabularx}
\newcolumntype{C}[1]{>{\centering\arraybackslash}p{#1}}

\usepackage{amsmath}
\usepackage{amsfonts}
\usepackage{amssymb}
\usepackage{dsfont}
\usepackage{pifont}
\usepackage{booktabs}
\usepackage{graphicx}
\usepackage{epstopdf}
\usepackage{epsfig}
\usepackage{framed}
\usepackage{makeidx}
\usepackage{siunitx}
\usepackage[capitalise]{cleveref}
\usepackage{hyperref}
\usepackage{placeins}
\usepackage[font=small,labelfont=bf]{caption}

\RequirePackage{csquotes}
\usepackage{bbm}

\begin{document}
\newgeometry{top=1.5cm,bottom=1.5cm,left=1.5cm,right=1.5cm,bindingoffset=0mm}

\vspace{-2.0cm}
\begin{flushright}
TIF-UNIMI-2025-10\\
Edinburgh 2024/12\\
CERN-TH-2025-106\\
\end{flushright}
\vspace{0.3cm}

\begin{center}
  {\Large \bf A Determination of $\alpha_s(m_Z)$ at aN$^3$LO$_{\bf QCD}\otimes {\bf NLO}_{\bf QED}$ Accuracy\\[0.2cm] from a Global PDF Analysis}
  \vspace{1.1cm}

{\bf The NNPDF Collaboration}: \\[0.1cm]
   Richard D. Ball$^1$,
   Andrea Barontini$^{2}$,
   Juan Cruz-Martinez$^3$,
   Stefano Forte$^2$,\\[0.1cm]
   Felix Hekhorn$^{4,5}$,
   Emanuele R. Nocera$^6$,
   Juan Rojo$^{7,8}$, and
   Roy Stegeman$^1$

    \vspace{0.7cm}

    {\it \small

    ~$^1$The Higgs Centre for Theoretical Physics, University of Edinburgh,\\
      JCMB, KB, Mayfield Rd, Edinburgh EH9 3JZ, Scotland\\[0.1cm]
    ~$^2$Tif Lab, Dipartimento di Fisica, Universit\`a di Milano and\\
      INFN, Sezione di Milano, Via Celoria 16, I-20133 Milano, Italy\\[0.1cm]
      ~$^3$CERN, Theoretical Physics Department, CH-1211 Geneva 23, Switzerland\\[0.1cm]
      ~$^4$University of Jyvaskyla, Department of Physics, P.O. Box 35, FI-40014 University of Jyvaskyla, Finland\\[0.1cm]
   ~$^5$Helsinki Institute of Physics, P.O. Box 64, FI-00014 University of Helsinki, Finland\\[0.1cm]
   ~$^6$ Dipartimento di Fisica, Universit\`a degli Studi di Torino and\\
      INFN, Sezione di Torino, Via Pietro Giuria 1, I-10125 Torino, Italy\\[0.1cm]
       ~$^7$Department of Physics and Astronomy, Vrije Universiteit, NL-1081 HV Amsterdam\\[0.1cm]
      ~$^8$Nikhef Theory Group, Science Park 105, 1098 XG Amsterdam, The Netherlands\\[0.1cm]

      }

\vspace{0.9cm}
{\bf \large Abstract}

\end{center}

We present a determination of the strong coupling $\alpha_s(m_Z)$ from a
global dataset including both fixed-target and collider data from deep-inelastic
scattering and a variety of hadronic processes, with a simultaneous
determination of parton distribution functions (PDFs) based on the NNPDF4.0 methodology.
This determination is performed at NNLO and approximate N$^3$LO
(aN$^3$LO) perturbative QCD accuracy, including  QED corrections and a
photon PDF up to NLO accuracy.
We extract $\alpha_s$ using two independent methodologies, both of which take into account the cross-correlation between $\alpha_s$ and the PDFs.
The two methodologies are validated by closure tests that allow us to
detect and remove or correct for several sources of bias, and  lead to
mutually consistent results.
We account for all correlated experimental uncertainties, as well as correlated theoretical uncertainties related to missing higher order perturbative corrections (MHOUs).
We study the perturbative convergence of our results and the impact of
QED corrections. We assess individual sources of uncertainty,
specifically MHOUs and the value of the top quark mass. We provide a
detailed appraisal of methodological choices, including the choice
of input dataset, the form of solution of evolution equation, the
treatment of the experimental covariance matrix, and the details of
Monte Carlo data generation.
We find $\alpha_s(m_Z)=0.1194^{+0.0007}_{-0.0014}$ at aN$^3$LO$_{\rm QCD}\otimes {\rm NLO}_{\rm QED}$ accuracy, consistent with the latest PDG average and with recent lattice results.

\clearpage

\tableofcontents

\section{Introduction}
\label{sec:introduction}

The precise knowledge of the strong coupling  $\alpha_s$ is one of the main bottlenecks towards reaching percent or
sub-percent accuracy in the computation of hadron collider
processes~\cite{Salam:2017qdl}.
Conversely, many of the most accurate
determinations of the strong coupling are obtained from processes that
involve hadrons in the initial
state~\cite{dEnterria:2022hzv,ParticleDataGroup:2024cfk}. These
determinations inevitably involve knowledge of  hadron structure,
as encoded in parton distribution functions (PDFs)~\cite{Gao:2017yyd,Kovarik:2019xvh}, and it has now
been known for some time~\cite{Forte:2020pyp} that  reliable unbiased
results can only be obtained if $\alpha_s$  and the PDFs are
simultaneously determined, as opposed to using a fixed PDF set.

Extractions of $\alpha_s(m_Z)$ together with the PDFs have been
carried out by various groups over the years, with the most precise
recent results obtained by groups that make use of a global dataset
involving several disparate pieces of experimental
information~\cite{Alekhin:2017kpj,Ball:2018iqk,Hou:2019efy,Cridge:2024exf}.
These PDF-based determinations differ in the input dataset, the accuracy of the
theory calculations, and the fitting methodology.
In particular, the NNPDF collaboration has presented several determinations at
NNLO QCD accuracy, based not only on increasingly wider datasets and
more refined PDF determination methodology, but also on a more
sophisticated treatment of the correlation between $\alpha_s$ and the
PDFs.

Specifically, in Ref.~\cite{Lionetti:2011pw} a first result based on
the NNPDF2.1~\cite{Ball:2011mu,Ball:2011uy} methodology was obtained,
by repeating  the PDF determination
for several fixed values of $\alpha_s$ and extracting $\alpha_s$ and
its uncertainty from the $\chi^2(\alpha_s)$ parabolic profile. While
this gives the correct central value, it generally underestimates the
uncertainty because the correlation between $\alpha_s$ and the PDFs
is not fully accounted for.  Indeed, for an accurate determination of the
uncertainty,  knowledge of the $\chi^2$ paraboloid in joint
$\alpha_s$ and PDF space is necessary. This in turn requires the
simultaneous determination of $\alpha_s$ and
the PDFs, as opposed to the determination of PDFs for different fixed
values of $\alpha_s$. This result was accomplished in
Ref.~\cite{Ball:2018iqk}, based on
NNPDF3.1~\cite{Ball:2017nwa} methodology,
through a correlated replica method (CRM), that involves performing a
family of PDF replica determinations using different $\alpha_s$ values
to each individual Monte Carlo data replica.

Since then, progress has been made in various directions: the   PDF
determination methodology, theory treatment and uncertainty treatment;
the $\alpha_s$ extraction methodology; and the validation methodology.
Concerning the PDF determination, the NNPDF3.1
methodology has been superseded by the more precise and
accurate NNPDF4.0~\cite{Ball:2021leu,NNPDF:2021uiq} methodology, based on modern
machine learning techniques. On the theory side, thanks to recent
progress on the N$^3$LO calculations of splitting functions
(see~\cite{Falcioni:2024xav,Falcioni:2024qpd,Kniehl:2025ttz} for the latest results)
it is now possible to determine PDFs at approximate N$^3$LO (aN$^3$LO) accuracy,
as done by MSHT~\cite{McGowan:2022nag} and
NNPDF~\cite{NNPDF:2024nan}. Moreover, it is now clear that the
inclusion of the photon PDF in joint QCD$\otimes$QED evolution
equations is necessary for percent accuracy, and  both MSHT~\cite{Cridge:2023ryv} and NNPDF~\cite{NNPDF:2024djq,Barontini:2024dyb} have
included QED effects in their aN$^3$LO PDF determinations (see also~\cite{MSHT:2024tdn} for their combination).

As far as uncertainties are concerned, it is now recognized that in order to
obtain accurate PDF
uncertainties  it is necessary to include correlated missing higher
order uncertainties (MHOUs) on the theory predictions for the
processes used  in the PDF determination. This was done
by MSHT at aN$^3$LO~\cite{McGowan:2022nag}  using a nuisance parameter
formalism, and by NNPDF both at NNLO~\cite{NNPDF:2024dpb} and
aN$^3$LO~\cite{NNPDF:2024nan} using the theory covariance
matrix formalism developed in
Refs.~\cite{Ball:2018lag,Ball:2018twp,NNPDF:2019vjt,NNPDF:2019ubu,Ball:2020xqw}
to account for MHOUs (as well as nuclear uncertainties).

As for the $\alpha_s$ extraction method, it was shown in
Ref.~\cite{Ball:2021icz} that the value and uncertainty on any theory
parameter, such as the parameters that determine the shape of the PDFs
or indeed $\alpha_s$, can be determined by Bayesian inference from
knowledge of the covariance matrix of theory parameters. When applied to $\alpha_s$ this theory covariance method (TCM)
provides an alternative way of extracting the strong coupling that
also fully keeps into account the correlation to PDFs.
Finally, it is now recognized that closure
tests~\cite{NNPDF:2014otw,Ball:2021leu,DelDebbio:2021whr,Harland-Lang:2024kvt,Barontini:2025lnl}
are necessary for a full validation of the methodology used
to determine PDF uncertainties, and thus also $\alpha_s$.

We present here a new determination of $\alpha_s$ that includes
all these developments. Specifically, we determine $\alpha_s$
and PDFs based on NNPDF4.0 methodology, using theory up to aN$^3$LO QCD,
with QCD$\otimes$QED including a photon PDF up to NLO, with full
inclusion of MHOUs.
Results are obtained using both the CRM and TCM.
These methodologies are validated by a closure
test that allows us to detect sources of
bias: specifically, those
related to the treatment of multiplicative
uncertainties and to positivity constraints.
We show that the two methodologies lead to consistent results, and we
appraise the impact of correlations between PDFs and $\alpha_s$ and
MHOUs. We assess perturbative convergence and the effect of QED
corrections and we study the impact of positivity.
We check the stability of our result upon a sizable number of
methodological and parametric variations, including the form of the solution of
evolution equations, the value of the top quark mass, and the Monte Carlo
data generation. We also study dataset dependence and specifically
show agreement with our previous $\alpha_s$ value of
Ref.~\cite{Ball:2021leu} if the same dataset is adopted.

The paper is organized as follows.
In Sect.~\ref{sec:review} we review the methods that we use for our
$\alpha_s$ determination: the CRM of Ref.~\cite{Ball:2018iqk}, and the
TCM of Ref.~\cite{Ball:2021icz} and its specific application to $\alpha_s$.
These two methodologies are validated by means of closure tests in
Sect.~\ref{sec:closure_tests}, where we detect and characterize possible sources of bias.
Our main results for $\alpha_s(m_Z)$ are presented in
Sect.~\ref{sec:results}, where we study their stability with respect
to variations of  theory, methodology,  and experimental input.
Conclusions are drawn in Sect.~\ref{sec:summary}, where we also
provide information on how to access our results (specifically
the  {\sc\small LHAPDF} grids~\cite{Buckley:2014ana}), all of which are made public.

\section{Methodologies for $\alpha_s(m_Z)$ extraction}
\label{sec:review}

We perform a simultaneous determination of $\alpha_s$ and PDFs
using two different methodologies that fully take into account the correlations
between $\alpha_s$ and the PDFs.
The first is the correlated replica method (CRM),
used for the extraction of $\alpha_s$ in Ref.~\cite{Ball:2018iqk}. It
is based
on a maximum likelihood estimate of all parameters which relies on
frequentist Monte Carlo resampling.
The second is based on the theory covariance method (TCM),
first introduced in Ref.~\cite{Ball:2021icz} as a means to
account for correlations between MHOUs on theory predictions obtained
from a given PDF set, and MHOUs on the predictions that had been used
to determine them.
The TCM, applied for the first time to the determination of $\alpha_s$ in
the present, relies on Bayesian posterior
parameter estimation.
The Monte Carlo method and the Bayesian method
are statistically equivalent for linear error propagation of Gaussian
uncertainties~\cite{Costantini:2024wby}, and below we will explicitly
check that indeed the TCM and CRM lead to consistent results both in a
closure test and with  real data.
It would be in principle also possible to determine $\alpha_s$
and the PDFs simultaneously using the SIMUnet
method~\cite{Iranipour:2022iak,Costantini:2024xae}, which involves interpolating the
theory for given PDF input as $\alpha_s$ is
varied, and then treating $\alpha_s$ as a parameter in the numerical
optimization; however, we do not use this technique in this paper.

\subsection{The correlated replica method}
\label{subsec:CRM}

For completeness and in order to set up the notation,
we provide a self-contained introduction to the CRM and a brief summary of the NNPDF methodology on which it is based,
referring the reader to Ref.~\cite{Ball:2018iqk} for more details.

The Monte Carlo inference method adopted by NNPDF is based on
starting from a Monte Carlo representation of the probability
distribution of the original experimental data, and determining for
each data replica an optimal PDF represented by a neural network,
found through conditional optimization of a
suitable loss function, thus obtaining
a Monte Carlo representation of the
probability distribution of PDF replicas.
The data replicas $\{ D_i^{(k)}\}$, where $i$ denotes the data points $i\in \{1,\ldots,N_{\rm dat}\}$ and $k$ the replica numbers $k\in\{1,\ldots,N_{\rm rep}\}$, are obtained by sampling the original data
from a multi-Gaussian distribution such that
\begin{equation}
  \lim_{N_{\rm rep}\to\infty} \mathrm{cov}\left(D_i^{(k)}, D_j^{(k)}\right)=C_{ij} \, ,
  \label{covmat}
\end{equation}
where $C$ is the total covariance matrix of the data, in
turn given by a sum of contributions of an experimental $t_0$
covariance matrix $C^{\rm exp}_{t_0}$~\cite{Ball:2009qv,Ball:2012wy} and a theory
covariance matrix $C^{\rm th}$ that includes MHOUs and other theory
uncertainties (such
as nuclear uncertainties)~\cite{Ball:2018lag,Ball:2018twp,NNPDF:2019vjt,NNPDF:2019ubu,Ball:2020xqw}:
\begin{equation}
  C_{ij} = C^{\rm exp}_{t_0,ij}+C^{\rm th}_{ij} \, .
  \label{eq:total_cov_mat}
\end{equation}

An optimal PDF replica, characterized by parameters ${\theta}^{(k)}$, is then determined for each data replica by
minimizing a loss function computed on a training subset of data and
stopping the training conditionally on the loss computed on the
remaining (validation) data subset. For the theoretical prediction of
the $i$-th data point evaluated with the $k$-th PDF replica, denoted $T_{i}({\theta}^{(k)},\alpha_s)$, the loss
function is
\begin{equation}
  \label{eq:chi2not}
  E^{(k)}\lp {\theta}^{(k)},\alpha_s\rp  =\frac{1}{ N_{\rm dat}} (T({\theta}^{(k)},\alpha_s) - D^{(k)})^T \, C^{-1} \,
  (T({\theta}^{(k)},\alpha_s) - D^{(k)} )\, ,
\end{equation}
where we have adopted a vector notation so that $T$ and $D$ are  $N_{\rm dat}$-component vectors, $T^T$ and $D^T$ the corresponding transpose vectors, and $C$ is an  $N_{\rm dat}\times N_{\rm dat}$ real symmetric matrix. All indices $i,j\in \{1,\ldots,N_{\rm dat}\}$ are then implicitly summed over, but the replica number $k$ and $\alpha_s$ dependence are left explicit.
Note that in the definition of Eq.~(\ref{eq:chi2not}) the theoretical
predictions, and thus the dependence on $\alpha_s$, enter both directly, as displayed, but also indirectly when constructing the $t_0$ covariance matrix $C^{\rm exp}_{t_0}$ and the theory covariance matrix $C^{\rm th}$.
As we will demonstrate in Sect.~\ref{sec:closure_tests}, this dependence of the covariance matrix on $\alpha_s$ may lead to biased results if not treated with care.

In the CRM, the PDFs parameters ${\theta}^{(k)}$ are determined for each fixed replica $k$, for a
number of different fixed values of
$\alpha_s$, thereby leading to an $\alpha_s$ dependent vector of
optimized parameters for each replica, that we denote
$\overline{\theta}^{(k)}(\alpha_s)$. A maximum
likelihood estimate of $\alpha_s$ for each data replica may then
be obtained as
\begin{equation}
  \alpha_s^{(k)} = \operatorname*{arg\,min}\left[E^{(k)}\left(\overline{\theta}^{(k)}(\alpha_s),\alpha_s\right)\right]\, .
\label{eq:alphacrm}
\end{equation}
A continuous function $E^{(k)}\left(\overline{\theta}^{(k)}(\alpha_s),\alpha_s\right)$
of $\alpha_s$ may be obtained by
interpolating the values of the loss
Eq.~(\ref{eq:chi2not}) obtained with all the given  values of $\alpha_s$,
which in
practice means fitting them to a parabola, or possibly a
higher order polynomial.
This then leads to a Monte Carlo representation of the probability
distribution in the joint $(\alpha_s$, PDF) space, whence the most likely value of $\alpha_s$ and associated confidence levels (CL)
can be determined,  as well as the correlations with the PDFs.
We refer to Ref.~\cite{Ball:2018iqk} for technical details on the implementation of the CRM in the NNPDF framework, which we follow in this work.

It is important to observe that the CRM provides a more reliable
estimate of the $\alpha_s$ uncertainty than that which is obtained by
neglecting the correlation between $\alpha_s$ and the PDF, as was
the case in the earlier NNPDF2.1-based determination of
$\alpha_s$~\cite{Lionetti:2011pw,Ball:2011us}. In this simpler
procedure (sometimes still
used today) the central
theory prediction
\begin{equation}\label{eq:tzero}
  T^{(0)}(\alpha_s) = \frac{1}{N_\mathrm{rep}} \sum_{k=1}^{N_\mathrm{rep}} T\left(\overline\theta^{(k)}(\alpha_s),\alpha_s\right),
\end{equation}
is used to evaluate the loss
\begin{equation}
  \label{eq:chi2not_simple}
  \chi^2(\alpha_s)  = ( T^{(0)}(\alpha_s)  - D)^T \,C^{-1} \,
  ( T^{(0)}(\alpha_s) - D )\, .
\end{equation}
The best fit value of $\alpha_s$ is then
\begin{equation}\label{eq:assimple}
  \alpha_s^{(\rm min)} = \operatorname*{arg\,min}\left[\chi^{2}(\alpha_s)\right]\, .
\end{equation}
with  68\% CL uncertainty found from the $\Delta\chi^2 = 1$ range about
the best-fit value.
This leads in general to an underestimate of the
uncertainty on $\alpha_s$, as demonstrated in
Ref.~\cite{Ball:2018iqk}, since it does not account for  the
correlation in uncertainty between $\alpha_s$ and the PDF.

\subsection{The theory covariance method}
\label{subsec:TCM}

The TCM was originally introduced in Ref.~\cite{Ball:2021icz} as a means to avoid double-counting of theory uncertainties
when computing predictions for a process that is correlated with data whose theoretical uncertainties have been included in the PDF determination.
However, the same technique can also be used to obtain a
Bayesian determination of the maximum likelihood value of any nuisance
parameter. Performing this determination for each data replica leads to
a determination of the probability distribution of the nuisance
parameter.
Hence, by viewing the deviation of the value of $\alpha_s$ (or indeed any other parameter entering the theory predictions) from its prior as a nuisance parameter, the method can be used to obtain a determination of $\alpha_s$ and its associated probability distribution.
Here, we first briefly summarize the general method, then describe its application to the determination of $\alpha_s$.

The starting observation is the well-known result that any correlated Gaussian uncertainty can be represented as a shift of
either the theory
or the data through a nuisance parameter $\lambda$:
$T \to T+\lambda \beta$ for a theory uncertainty, or $D \to D-\lambda \beta$ for a data
uncertainty.
Henceforth we will assume for definiteness that the
correlated uncertainty is a theory uncertainty (hence the choice of
sign in the definition of $\lambda$) though the treatment is entirely
symmetric under the interchange of $T$ and $D$.
Such a correlated shift of the theory predictions is equivalent to adding a contribution
\begin{equation}\label{eq:cbeta}
  S_{ij}=\beta_i\beta_j \, ,
\end{equation}
to the covariance matrix $C_{ij}$, Eq.~(\ref{eq:total_cov_mat}).

In a Bayesian framework, this can be easily proven as follows. The
probability $P(T|D)$ can be obtained by marginalizing over $\lambda$ the joint
probability  $P(T|D,\lambda)$ multiplied by the prior $P(\lambda)$:
\begin{equation}\label{eq:bayesian}
  P(T|D)=\int d\lambda P(T|D,\lambda)P(\lambda).
  \end{equation}
For Gaussian distributed observables
\begin{equation}
  \label{eq:chi2not_bayesian_lam}
  P(T|D,\lambda) \propto \exp\lc -\frac{1}{2} (T + \lambda\beta- D)^T C^{-1} \,
    (T+\lambda\beta- D) \rc \, ,
\end{equation}
so, if the prior $P(\lambda)$ is a univariate Gaussian centered at zero
\begin{equation}\label{eq:plambda}
  P(\lambda)\propto \exp \lp  -\frac{\lambda^2}{2}\rp \, ,
\end{equation}
then it is straightforward to perform the integral over
$\lambda$ by completing the square, to give
\begin{equation}
  \label{eq:chi2not_bayesian}
  P(T|D) \propto \exp\lc -\frac{1}{2} (T- D)^T (C+S)^{-1} \,
  (T- D) \rc.
\end{equation}
Thus the correlated uncertainty parametrized by the nuisance parameter $\lambda$ may be incorporated simply by adding the contribution $S$, given by
Eq.~(\ref{eq:cbeta}), to the original covariance matrix $C$, Eq.~(\ref{eq:total_cov_mat}).

The advantage of this point of view is that  Bayes' theorem also
determines the posterior distribution of the nuisance parameter:
\begin{equation}\label{eq:Plam}
P(\lambda | T,D) \propto \exp\lp -\frac{1}{2}Z^{-1}\lp \lambda - \bar{\lambda}(T,D)\rp^2\rp \, ,
\end{equation}
which is a  Gaussian of width $Z$ centered at
$\bar{\lambda}$, with
\begin{align}
\label{eq:lambar}
\bar{\lambda}(T,D)&=\beta^T (C+S)^{-1} (D-T) ,\\
\label{eq:lamwidth}
Z &= 1-\beta^T (C+S)^{-1} \beta .
\end{align}

This way of treating correlated uncertainties can be integrated within
the NNPDF methodology by simply  representing any of the correlated
uncertainties included in the covariance matrix  through a nuisance
parameter. Assuming for the sake of argument that we start from some
covariance matrix $C$,
Eq.~(\ref{eq:total_cov_mat}), and
we want to add to $C$ a new correlated theory uncertainty, represented by
a covariance matrix $S$, and uncorrelated to any of the uncertainties
already included in $C$, we perform
determinations of $\bar{\lambda}$ by generating Monte Carlo data replicas as in
Eq.~(\ref{covmat}), then fit them using the loss function
Eq.~(\ref{eq:chi2not}), but in each case with the covariance matrix
$C$ replaced by  $C+S$. In this way for each data replica  $D^{(k)}$ we
obtain
associated PDF parameters $\overline{\theta}^{(k)}$, and thus theoretical predictions  $T^{(k)}$, which we can use to determine an ensemble of replicas $\bar{\lambda}^{(k)}$ of the nuisance parameters, with expectation value
\begin{equation}
\label{eq:mean_nuisance_parameter}
\bar{\lambda}^{(0)} =
\beta^T  (C+S)^{-1}(D- T^{(0)}) \, ,
\end{equation}
where $T^{(0)}$ is the central prediction, Eq.~(\ref{eq:tzero}), and $D$
the central data point. The variance  of the nuisance parameter over the replica sample can also be computed analytically:
\begin{equation}
\label{eq:cov_nuisance_parameter}
\bar{Z}\equiv 1 - \beta^T (C+S)^{-1}\beta
+\beta^T (C+S)^{-1} X (C+S)^{-1}\beta  \,,
\end{equation}
where  $X$ is the covariance matrix of the theoretical predictions, averaged over the $N_{\rm rep}$ PDF replicas:
\begin{equation}
\label{eq:matrix_X_pdfunc}
X_{ij}=  \frac{1}{N_\mathrm{rep}}\sum_{k=1}^{N_\mathrm{rep}}( T_i^{(k)} - T_i^{(0)})(T_j^{(k)} - T_j^{(0)})  \, .
\end{equation}
The whole derivation can be straightforwardly extended to the case of
multiple nuisance parameters and we refer to Ref.~\cite{Ball:2021icz} for further details.

We may apply this procedure to the determination of $\alpha_s(m_Z)$ by
simply viewing the deviation of $\alpha_s$, that was hitherto kept
fixed  at $\alpha_s(m_Z)=\alpha_s^{0}$, as a nuisance parameter. The
value $\alpha_s^{0}$ is then viewed as a prior, the nuisance parameter
is
\begin{equation}
\lambda = \alpha_s - \alpha_s^{0}\, .
\end{equation}
and  the posterior
distribution and uncertainty of $\lambda$ are determined from the data.
Taking $\alpha_s^\pm$ either side of this central value to establish
a prior uncertainty, the nuisance parameter is endowed with  a Gaussian prior
centered on zero, with width $\Delta\alpha_s^\pm = \alpha_s^\pm - \alpha_s^{0}$. We assume a symmetric interval, so
$|\Delta\alpha_s^+|=|\Delta\alpha_s^-|=\Delta\alpha_s$. The value of
$\Delta\alpha_s$ fixes the width of the prior, which should be chosen
wide enough that final results are independent of the prior, though
not so wide that one can no longer use standard linear error
propagation.

The vector of prior widths of theory predictions $\beta$ is
determined by linearizing the dependence of the central theoretical predictions $T^{(0)}(\alpha_s)$, Eq.~(\ref{eq:tzero}), around $\alpha_s^{0}$:
\begin{equation}
\label{eq:taylor_theory_predictions}
T^{(0)}\lp\ \alpha_s\rp =  T^{(0)}(\alpha_s^{0}) + (\alpha_s -
\alpha_s^{0}) \frac{\partial T^{(0)}\lp \alpha_s\rp }{\partial
  \alpha_s}\Bigg|_{\alpha_s=\alpha_s^{0}}+O((\Delta\alpha_s)^2)=
T^{(0)}(\alpha_s^{0}) + \lambda\beta,
\end{equation}
with
\begin{equation}\label{eq:betadef}
\beta = \frac{\partial T^{(0)}\lp \alpha_s\rp }{\partial \alpha_s}\Bigg|_{\alpha_s=\alpha_s^{0}}= \frac{\Delta T^\pm}{\Delta\alpha_s}+O((\Delta\alpha_s)^2),
\end{equation}
where
\begin{equation}
\Delta T^\pm = T^{(0)}(\alpha_s^\pm) - T^{(0)}(\alpha_s^{0}).
\end{equation}
Since the dependence of the theory predictions on $\alpha_s$ is slightly non-linear, we approximate the theory covariance matrix used in the fit by averaging over the positive and negative variations:
\begin{equation}
  S_{ij} = \beta_{i}\beta_{j}(\Delta\alpha_s)^2 = \frac{1}{2} \lp \Delta T_i^+\Delta T_j^+ +\Delta T_i^-\Delta T_j^- \rp \, .
\end{equation}
We have checked that our results are independent of this prescription,
provided $\Delta\alpha_s$ is sufficiently small: in practice
just a few percent of $\alpha_s$, so that corrections are of
order tenths permille.

Once we have determined $S$, we simply perform a single PDF determination in
which generation of data replicas and fitting using the loss function
are both performed using the combined covariance matrix $C+S$. The
additional theory covariance matrix $S$ included in this fit allows
the PDFs to accommodate the prior uncertainty in $\alpha_s$, and
evaluating the nuisance parameters replica by replica gives us the
posterior distribution. Specifically, using
Eqs.~(\ref{eq:mean_nuisance_parameter})
and~(\ref{eq:cov_nuisance_parameter}), the best-fit value is
\begin{equation}
\label{eq:TCM_final_1}
\bar{\alpha}_s = \alpha_s^{0}  +\overline{\lambda}^{(0)} =
\alpha_s^{0} + \beta^T  \left(C+S\right)^{-1} \lp D- T^{(0)} \rp \Delta\alpha_s \, ,
\end{equation}
while the associated standard deviation is given by
\begin{equation}
\label{eq:TCM_final_2}
\sigma_\alpha = \lp 1 - \beta^T \left(C+S\right)^{-1}\beta
+\beta^T \left(C+S\right)^{-1} X \left(C+S\right)^{-1}\beta\rp^{1/2} \Delta\alpha_s\, .
\end{equation}

Independence of the final value of  the choice of  prior can be achieved
by iteration of the whole procedure, with the central value $\alpha_s^{0}$ of the new prior taken as the posterior value $\bar{\alpha}_s$ of the previous fit.

\subsection{Settings for $\alpha_s$ extraction}
\label{subsec:comp}

We use both the CRM and TCM for $\alpha_s$ determination,
and show that they lead to consistent results.
For the CRM, we  determine correlated replicas for values of the
strong coupling in the range $\alpha_s(m_Z)\in [0.114, 0.125]$ with increments
of $\Delta \alpha_s=0.001$. For the TCM we  choose $\Delta \alpha_s=0.002$
for the prior while the central value of the prior is updated iteratively until
the prior and posterior agree. Prior independence is also
explicitly checked.
It is clear that the TCM is generally, and
especially with these choices, computationally more efficient than the
CRM, since once the theory covariance matrix is computed it requires only a single fit and the evaluation of the formulae Eqs.~(\ref{eq:TCM_final_1},\ref{eq:TCM_final_2}), as opposed to performing a large
number of correlated fits equal to the set of chosen discrete $\alpha_s$ values in the CRM. Moreover, the TCM only requires the determination of PDFs for a value of $\alpha_s$ close to the physical value, while the CRM requires PDF determinations over a wider range of values, some of which are quite far from the physical value. This may be significant, since the hyperoptimization of the fit parameters is performed only for a single value of $\alpha_s$ ($0.118$ in NNPDF4.0).

\section{Closure tests}
\label{sec:closure_tests}

Closure tests~\cite{DelDebbio:2021whr,Barontini:2025lnl} as a means to validate PDF sets were first introduced for the NNPDF3.0 determination~\cite{NNPDF:2014otw},
subsequently used for NNPDF4.0~\cite{Ball:2021leu} and recently adopted by other groups~\cite{Harland-Lang:2024kvt}.
Here we use them, for the first time, to validate an $\alpha_s$
determination: we generate ``synthetic'' data with a known
underlying true value of $\alpha_s$ and the PDFs, and we then show
that the value
of $\alpha_s$
obtained by running our methodology, blind to this underlying value,
agrees with it.
This provides us with an extremely
stringent test, because in the context of a closure
test it is possible to regenerate synthetic data $N_r$ times, corresponding
to $N_r$ ``runs of the universe''. It is then possible not only to
check that the values of $\alpha_s$ obtained in each run of the
universe are distributed about the true value according to their
nominal uncertainty $\sigma_\alpha$, but also that the mean over runs
of the universe
agrees with the true value with a smaller uncertainty
$\sigma_\alpha/\sqrt{N_r}$. In fact, the closure test has allowed us
to detect two possible sources of bias: first, related to the treatment of
multiplicative uncertainties, and second, to positivity constraints.

\subsection{Closure testing methodology and settings}
\label{subsec:settings}

\paragraph{Methodology.}
Closure tests are performed by generating synthetic data
(referred to as L1 data) according to a known underlying law: in our
case a known set of PDFs and value of $\alpha_s$. The true values are
referred to as L0 data, and the L1 data are distributed about them
according to the full experimental covariance matrix.
The NNPDF methodology is then run on these synthetic data: this in particular involves
generating Monte Carlo data replicas, referred to as L2 data.
The whole procedure is
repeated $N_r$ times, corresponding to $N_r$ independent runs of the
universe. Hence, $N_r$ sets of L1 data are generated, from each of which
we extract a value of $\alpha_s$ using the CRM and the TCM, as
discussed in Sect.~\ref{sec:review}. For
reasons of computational cost (see Sect.~\ref{subsec:comp}) the CRM is run for a smaller
number $N_r$ of L1 runs than the TCM.

The accuracy of the results for $\alpha_s$ obtained by each methodology can be
tested using the bias-variance ratio or mean normalized
bias~\cite{DelDebbio:2021whr,Barontini:2025lnl}:
\begin{equation}
\label{eq:alphas_Rbv}
\mathcal{R}_{\rm bv} =  \sqrt{\frac{1}{N_r}  \sum_{j=1}^{N_r} \left(\mathcal{R}_{\rm
    bv}^{(j)}\right)^2},\end{equation}
with
\begin{equation}\label{eq:alphas_Rbvi}
  \quad \mathcal{R}_{\rm  bv}^{(j)}=  \frac{\alpha_s^{(j)} - \bar{\alpha}_s}{\sigma_\alpha^{(j)}},
\end{equation}
where $\bar\alpha_s$ is the true underlying value of $\alpha_s$, and
$\bar{\alpha}_s^{(j)}$ and $\sigma_\alpha^{(j)}$ are respectively the
central value and uncertainty on $\alpha_s$ obtained in the $j$-th run
of the universe and the sum runs over the $N_r$ runs of the
universe. The bias-variance ratio of
Ref.~\cite{DelDebbio:2021whr}, and the normalized bias of
Ref.~\cite{Barontini:2025lnl}, when considering several
correlated quantities differ in the treatment of correlations, but
coincide  when considering a single quantity as in our case.
The
normalized bias $\mathcal{R}_{\rm bv}^{(j)}$ should follow a univariate normal distribution, of
which the bias-variance ratio is the  variance, and should thus
equal one for perfectly faithful uncertainties.
The uncertainty on $\mathcal{R}_{\rm bv}$ can be estimated via the bootstrap method (see e.g.\ Ref.~\cite{Barontini:2025lnl}).

\paragraph{Settings.}
We generate data selecting as true
underlying value
\begin{equation}
    \bar{\alpha}_s(m_Z) = 0.118 \, ,
\end{equation}
and the central value of a (QCD-only) NLO PDF as the true underlying PDF.
Since the closure test assumes that  the data exactly reproduces the predictions,
the particular choice of underlying theory is immaterial and there are
no missing higher order contributions and associate uncertainties.

We perform $N_r=25$ determinations of $\alpha_s$ using the  CRM, each
based on a set of  $N_{\text{rep}} = 250$ replicas for each of the 12
values of $\alpha_s$ under consideration (see
Sect.~\ref{subsec:comp}). With the TCM
we perform
$N_r=100$ determinations, each based on a set of
$N_{\text{rep}} = 550$ replicas. These numbers are before the post-fit selection to filter outliers~\cite{Ball:2021leu}.

\subsection{Results}
\label{subsec:ctresults}

\begin{table}[t]
  \centering
  \renewcommand{\arraystretch}{1.4}
\small
\begin{spacing}{1.3}
\begin{tabularx}{\textwidth}{XXcccc}
  \toprule
  Method & Settings
  &  $\langle \alpha_s(m_Z)\rangle $
  &  $\langle \sigma_\alpha\rangle /\sqrt{N_r}$
  & pull $P$
  &  $\mathcal{R}_{\rm bv}$ \\
  \midrule
  CRM &$C(\alpha_s)$
  & $0.119450$ & $0.000077$ & 19
  & $3.8 \pm 0.16$ \\
    \midrule
  CRM &fixed $C$
  & $0.118152$ & $0.000070$ & 2.2
  & $0.97 \pm 0.11$ \\
  TCM & fixed $C$
  & $0.118132$ & $0.000039$ & 3.4
  & $0.80 \pm 0.06$ \\
  \midrule
  CRM & fixed $C$, no positivity
  & $0.118029$ & $0.000077$ & 0.38
  & $0.80 \pm 0.09$ \\
  TCM & fixed $C$, no positivity
  & $0.117984$ & $0.000041$ & 0.39
  & $0.71 \pm 0.05$ \\
\bottomrule
\end{tabularx}
\end{spacing}

    \vspace{0.2cm}
  \caption{Results of the closure test for the $\alpha_s$
    determination performed with the CRM and TCM with different
    settings. From top to bottom, we show results obtained with a covariance matrix that
    depends on $\alpha_s$, with fixed covariance matrix, and without
    positivity (see text). In each case we
    show the mean $\langle \alpha_s\rangle$ Eq.~(\ref{eq:almean}),
    the uncertainty of the mean, which is by a factor
    $\sqrt{N_r}$ smaller than the mean uncertainty $\sigma_\alpha$
    Eq.~(\ref{eq:sigmasmean})
    on the
    value found in each run, the pull Eq.~(\ref{eq:pulldef}) and the
    bias-variance ratio Eq.~(\ref{eq:alphas_Rbv}), with uncertainty estimated via the bootstrap method.
}
  \label{tab:Rbv_alphas}
\end{table}

\paragraph{Methodological choices.}
We have considered a large number of possible methodological choices
and variations, both concerning the NNPDF methodology in general, and the
$\alpha_s$ determination in particular, in order to assess whether
any of them would affect the faithfulness of the $\alpha_s$ value.
Specifically, with the CRM in the determination of the best-fit
$\alpha_s$, Eq.~(\ref{eq:alphacrm}), we interpolated the available discrete
values of $E^{(k)}(\overline{\theta}^{(k)}(\alpha_s),\alpha_s)$ with
polynomials of increasingly higher order; we used  $\ln \alpha_s$
instead of $\alpha_s$ as a variable; we checked the effect of
following the multi-batch procedure of Ref.~\cite{Ball:2018iqk} in
which each data replica is fitted several times and the best fit is
selected vs.\ a single-batch. None of these
variations had any significant effect~\cite{barontiniphd}.
In the TCM, we significantly increased the width of the prior,
with no visible effect.
For both CRM and TCM we also generated L2 data using either
the experimental covariance matrix or the $t_0$ covariance matrix  (see
Ref.~\cite{Ball:2014uwa}, specifically Table~9, for a discussion of
the difference between the two); Again, this variation did not have any significant effect.

However, we did find two methodological choices that do have an impact on
the determination of $\alpha_s$, namely the treatment of multiplicative uncertainties in the experimental covariance matrix and the treatment of positivity. We discuss each of them in turn.

\paragraph{Multiplicative Uncertainties}

Both the experimental and theoretical covariance matrix $C^{\rm exp}_{t_0}$ and $C^{\rm th}$ Eq.~(\ref{eq:total_cov_mat}) depend on
the value of $\alpha_s$. Indeed, the  $t_0$ experimental covariance
matrix $C^{\rm exp}_{t_0}$~\cite{Ball:2009qv} is computed using the theory predictions from a previous
fit to determine multiplicative uncertainties, and the theory
covariance matrix $C^{\rm th}$ is found performing scale variations,
whose size is manifestly dependent on the value of $\alpha_s$.

It must consequently be decided whether, when varying the value of
$\alpha_s$ in the theory prediction used  to determine its
best-fit, the
value of $\alpha_s$ in the computation of the covariance matrix
should also be varied, or not. In the closure test, of
course, as there is no MHOU, only the effect of this choice for
$C^{\rm exp}_{t_0}$ is relevant.
The test is most easily performed in
the CRM, where the theory predictions are computed for a fixed set
of value of $\alpha_s$, and a loss
$E^{(k)}(\overline{\theta}^{(k)}(\alpha_s),\alpha_s)$
Eq.~(\ref{eq:alphacrm}) is then determined for each value.  The question is then whether the same
covariance matrix should be used when computing the loss for each
value of $\alpha_s$, or whether the covariance matrix should be
re-determined for each value of $\alpha_s$ along with the theory
prediction.

The value of $\alpha_s$ obtained when varying the covariance
matrix as a function of $\alpha_s$ is shown in the first row of
Table~\ref{tab:Rbv_alphas}. We display there the weighted mean over the $N_r$ runs
\begin{equation}\label{eq:almean}
\langle \alpha_s(m_Z) \rangle=\frac{\sum_{j=1}^{N_r}
\frac{\alpha^{(j)}_s(m_Z)}{\left(\sigma^{(j)}_\alpha\right)^2}}{\sum_{j=1}^{N_r}\frac{1}{\left(\sigma^{(j)}_\alpha\right)^2}}\,,
\end{equation}
where the weighted uncertainty is
\begin{equation}\label{eq:sigmasmean}
\langle \sigma_\alpha\rangle= \frac{1}{\sqrt{\sum_{j=1}^{N_r}
\frac{1}{\left(\sigma^{(j)}_\alpha\right)^2}}}\, .
\end{equation}
In the same table we also show the uncertainty of the mean, given by
$\langle \sigma_\alpha\rangle/\sqrt{N_r}$, the pull
\begin{equation}\label{eq:pulldef}
  P=\frac{\frac{1}{N_r}\sum_{j=1}^{N_r}\lp \alpha_s^{(j)} - \bar{\alpha}_s\rp}{\langle\sigma_\alpha\rangle/N_r},
\end{equation}
and the bias-variance ratio.
It is clear that the closure test fails: the bias-variance ratio shows that the deviation of results from truth is on average four times bigger than the nominal
uncertainty. Note that the pull is correspondingly
$P\approx\sqrt{N_r}\mathcal{R}_{\rm bv}\approx 20$.

The value of $\alpha_s$ extracted when keeping the covariance matrix
fixed, shown in the second row of Table~\ref{tab:Rbv_alphas}, is instead free of this
problem. The bias-variance ratio is now somewhat smaller than one,
indicating that the mean-square deviation of $\alpha_s$ is
consistent with its stated uncertainty, with, in fact, a
slight uncertainty overestimation. This agrees with the
result found in Ref.~\cite{Barontini:2025lnl} for PDFs. The value of
$\alpha_s$  found using the TCM, where the covariance matrix is kept
fixed by construction since only the dependence on $\alpha_s$ through
the theory predictions is included in Eq.~(\ref{eq:taylor_theory_predictions}), is given in the third row of the table,
and it is also in agreement, with a bias-variance ratio less than one.
We have also checked that the same consistent result
is reproduced if all uncertainties are treated as additive. Indeed, in
this case the $t_0$ matrix is not used at all, so the covariance matrix
becomes independent of $\alpha_s$.

This somewhat counter-intuitive result can be explained by noting that
recomputing the covariance matrix as a function of $\alpha_s$
introduces a dependence of the experimental correlated
systematics on $\alpha_s$. Since many hadronic cross-sections increase
as $\alpha_s$ increases,  this then tends to
make multiplicative uncertainties larger for larger $\alpha_s$,
and thus the loss smaller, thereby leading to
an upward bias in the best-fit value. We conclude that a consistent
determination of $\alpha_s$ requires keeping the covariance matrix
fixed as $\alpha_s$ is varied, a result that would have been difficult
to establish without the closure test.

\paragraph{Positivity}

Closer inspection of Table~\ref{tab:Rbv_alphas} reveals that
while the value of $\alpha_s$  determined with fixed covariance matrix
deviates on average from the true value by an amount which is
consistent with its nominal uncertainty, it nevertheless displays a
pull $P$ well above one. We have checked that this pull remains
approximately constant when increasing the number $N_r$ of runs: the
deviation of  the mean from the truth decreases but
so does its  uncertainty. This means that whereas in each run the
deviation of the best-fit $\alpha_s$ from truth is consistent with its
uncertainty (because $R_{bv}\sim 1$), the distribution of best-fits about the true value is
asymmetric, so on average biased. In a closure test this bias can be
reduced by increasing the number $N_r$ of runs, but a real-world
determination of course will consist of a single run, hence it is
important to understand the origin of the bias.

\begin{figure}[t]
    \centering
    \includegraphics[width=0.475\linewidth]{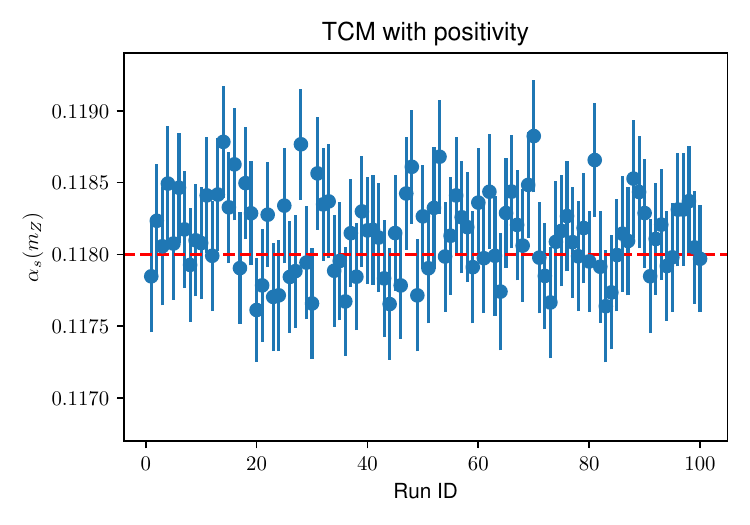}
    \includegraphics[width=0.486\linewidth]{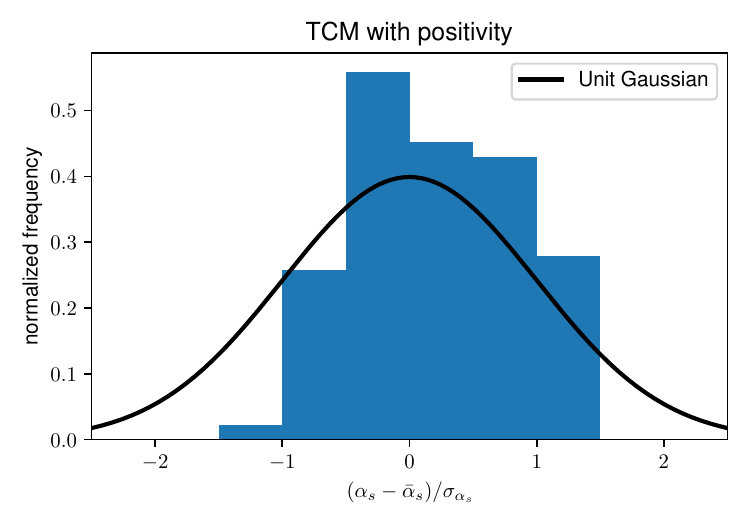}\\
    \includegraphics[width=0.475\linewidth]{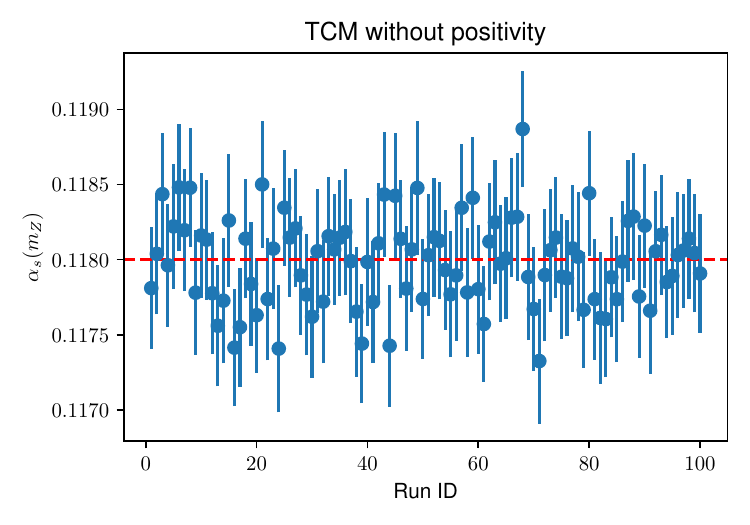}
    \includegraphics[width=0.486\linewidth]{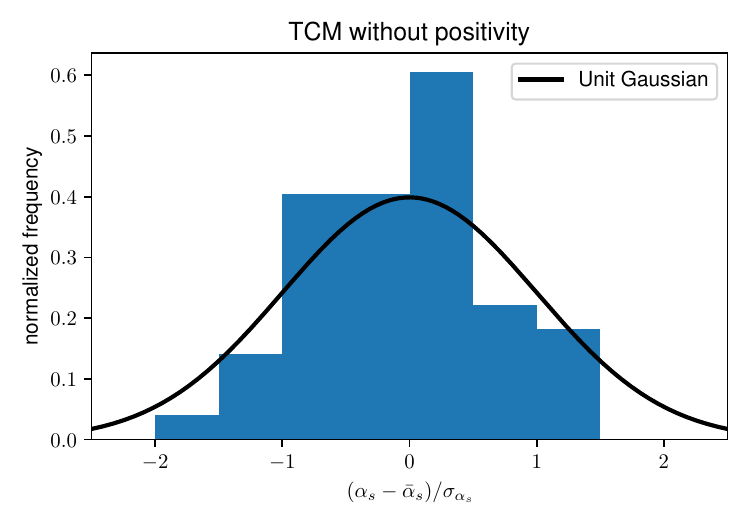}
    \caption{The best-fit values of $\alpha_{s}^{(j)}$ and the
      associated one standard deviation uncertainties
      $\sigma^{(j)}_\alpha$ obtained using the TCM in the
      $N_r=100$ individual runs of the
      closure tests (left), and the corresponding distribution
      of normalized bias $\mathcal{R}_{\rm
    bv}^{(j)}$ Eq.~(\ref{eq:alphas_Rbvi}) (right).     For reference,
      a univariate zero-mean Gaussian is also displayed in the right
      panels.
Results obtained both
      when imposing positivity (top) and when not
      imposing it (bottom) are shown.}
    \label{fig:failed_multiCT}
\end{figure}

We have traced this bias to the positivity constraints imposed in the PDF fit,
see Sect.~3.1.3 of Ref.~\cite{Ball:2021leu} for a detailed discussion.
Results found both with the TCM and CRM after removing these constraints are
shown in Table~\ref{tab:Rbv_alphas}. It is clear that while the bias-variance
ratio is unchanged, the pull is now below one, showing that the bias has
disappeared.
Also,  in Fig.~\ref{fig:failed_multiCT} we compare the results
obtained with
and without positivity for the
individual $N_r$ determinations of $\alpha_s(m_Z)$ using  the
TCM.
We show both a comparison of the result with its central value
$\alpha_s^{(j)}$ and uncertainty $\sigma_\alpha^{(j)}$ to the
underlying truth $\bar{\alpha}_s=0.118$, and the histogram of
normalized bias $\mathcal{R}_{\rm bv}^{(j)}$,
Eq.~(\ref{eq:alphas_Rbv}), superposed to a univariate Gaussian, which
is its expected distribution.

It is clear from the figure that the
distribution of $\alpha_s^{(j)}$ values is to the same good
approximation Gaussian with or without positivity, in agreement with
the fact that the bias-variance ratio with and without positivity 
remains the same: the distribution of results about the mean is in
each case compatible with its uncertainty and symmetric. However, the
distribution of results about the true value without positivity is
also symmetric, while with positivity it is
biased, as it is clear from Fig.~\ref{fig:failed_multiCT} (left) where
it is clear that with positivity the number of values of
$\alpha_s^{(j)}$ above the horizontal line is larger than the number
of values below. We conclude that positivity results in a bias that
produces an offset of the center of the distribution of
$\alpha_s^{(j)}$  values with respect to
the true value.

The impact of  positivity can be understood by noting  that in the vicinity
of kinematic boundaries the data uncertainty is necessarily non-Gaussian,
because a Gaussian always has an infinite tail which extends in the region of
negative cross-sections. However, experimental data uncertainties are assumed to
be Gaussian and treated as such in the data replica generation, which
may generate negative data replicas. This  may then lead to an
inconsistency between the distribution of optimized PDF replicas,
which are constrained to lead to positive predictions,
and that of the underlying data.

Tracing which datasets lead to the effect is however difficult, since
all data are correlated through their PDF dependence, PDFs are in turn
correlated by the momentum sum rule, and there might be an interplay
between data, and theoretical positivity constraints that are also
imposed in the fit~\cite{Ball:2021leu} in order to ensure that physical
observables remain positive even outside the data region.

\begin{figure}[t]
    \centering
    \includegraphics[width=0.475\linewidth]{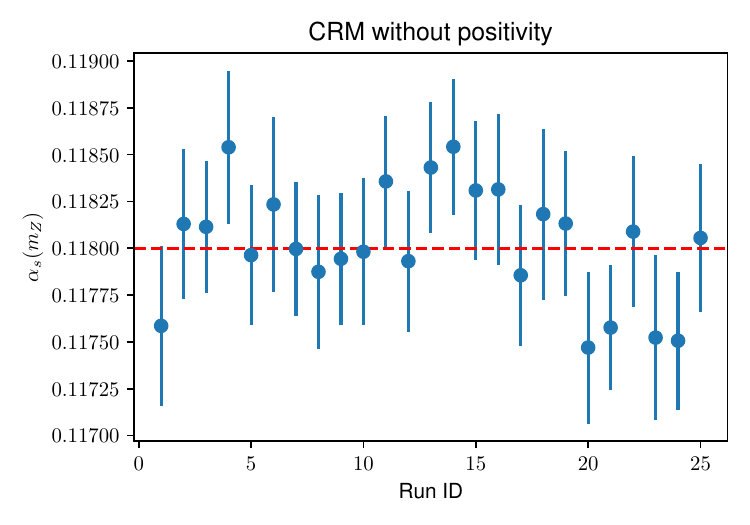}
    \includegraphics[width=0.486\linewidth]{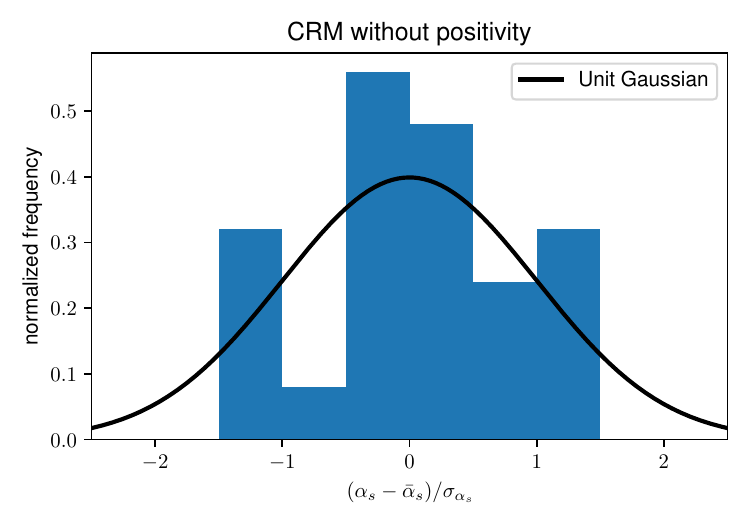}
    \caption{Same as Fig.~\ref{fig:failed_multiCT} (bottom) but now
      for results obtained in  the $N_r=25$ runs of the CRM.
    }
    \label{fig:alphas_comparison_multiCT}
\end{figure}

\paragraph{Closure Results.}
The two determinations of $\alpha_s$ shown in Table~\ref{tab:Rbv_alphas} after removing positivity constraints (bottom two entries)
satisfy the closure test. Note that the central values of the two determinations are not exactly the same  since the CRM result is determined from $N_r=25$ runs  and the TCM result from $N_r=100$ runs.
However, the average uncertainty $\langle \sigma_\alpha\rangle$ Eq.~(\ref{eq:sigmasmean}) agrees: the two determinations have the same pull, well below one.
The agreement of results found with the two methods is also
demonstrated by repeating the 
TCM plot of  Fig.~\ref{fig:failed_multiCT}, but now using the CRM, see
Fig.~\ref{fig:alphas_comparison_multiCT}

Note also that the uncertainty on the CRM prediction is determined for each run from the variance of the set of $N_{\rm rep}$ $\alpha_s^{(k)}$ values Eq.~(\ref{eq:alphacrm}), while the uncertainty
on the TCM prediction is found using the analytic formula
Eq.~(\ref{eq:TCM_final_2}). We have checked that the result for the
TCM uncertainty found using the analytic formula agrees with that
computed from the standard deviation
over the replica sample, and that the 68\% CL over replicas differs only at the
sub-permille level.
This indicates that the distribution of $\alpha_s$ found using the TCM is Gaussian to very good approximation, as expected.

Finally, we test for independence of the prior of results obtained using the TCM. This is especially important in view of the fact that in the previous tests we always used a prior centered at the true value. To this purpose, we have determined the posterior value of $\alpha_s$ by taking as a prior $\alpha_s(m_Z)=0.117$ or  $\alpha_s(m_Z)=0.119$.
We respectively find $\alpha_s(m_Z)=0.11801$ and
$\alpha_s(m_Z)=0.11811$ as posterior values. This proves that the
method converges rapidly:  the subsequent iteration would then
essentially coincide with our previous TCM determination, thereby
confirming prior independence. Note that in all cases, as discussed in Sect.~\ref{subsec:TCM} the width of the
prior is $\Delta\alpha_s=0.002$, hence it is much wider than the
positivity bias, i.e. the shift of the best-fit central value due to
positivity.

We conclude that the closure test is successful: the CRM and the TCM
lead to results in agreement with each other, unbiased and with
faithful uncertainties. We further conclude that unbiased results are
obtained with a covariance matrix that does not vary with $\alpha_s$,
and in the absence of positivity constraints. In the presence of
positivity constraints, results are still Gaussianly distributed, but
converge to 
biased result, offset by a positive amount with respect to the true value.

\section{The strong coupling at aN$^3$LO accuracy}
\label{sec:results}%

We now present the main result of this work, namely the extraction of $\alpha_s(m_Z)$ up to  aN$^3$LO QCD and NLO QED accuracy.
First we present our baseline determination, discussing its methodological aspects and perturbative stability and providing our assessment of its total uncertainty.
We then check that our error estimate is robust, by verifying that our result is stable upon various possible methodological variations.
We finally compare our result to previous determinations both by us and by other groups.


\begin{table}[t]
  \centering
  \renewcommand{\arraystretch}{1.4}
\small
\begin{spacing}{1.3}
\begin{tabularx}{\textwidth}{Xccc}
  \toprule
Perturbative order 
    & TCM  & CRM & best value\\
  \midrule
  
NNLO$_{\rm QCD}$ &
$0.1198 \pm 0.0008$
&
$0.1199 \pm 0.0006$ & $0.1198^{+0.0007}_{-0.0010}$\\

NNLO$_{\rm QCD}\otimes$NLO$_{\rm QED}$
&
$0.1203 \pm 0.0007$
&
$0.1201 \pm 0.0006$ & $0.1203^{+0.0007}_{-0.0010}$\\
aN$^3$LO$_{\rm QCD}$ 
&
$0.1192 \pm 0.0007$
&
$0.1191 \pm 0.0008$  & $0.1192^{+0.0007}_{-0.0013}$
\\
aN$^3$LO$_{\rm QCD}\otimes$NLO$_{\rm QED}$
&
$0.1194 \pm 0.0007$
&
$0.1194 \pm 0.0007$  & $0.1194^{+0.0007}_{-0.0014}$\\
\bottomrule
\end{tabularx}
\end{spacing}

    \vspace{0.2cm}
    \caption{Baseline results for  $\alpha_s(m_Z)$ obtained at
      different perturbative orders, using the CRM and TCM
      respectively discussed in Sects.~\ref{subsec:CRM} and
      \ref{subsec:TCM}. The final column gives our best value at each
      perturbative order, obtained using the TCM result.
      The asymmetric uncertainty is obtained by accounting for the systematic uncertainty associated to the
      positivity bias (see text).
}
  \label{tab:alphas_baseline}
\end{table}

\subsection{Baseline results}
\label{sec:baseline}

Our baseline results are obtained using the same NNPDF4.0 dataset, methodology and theory predictions
as in Refs.~\cite{NNPDF:2024dpb,NNPDF:2024nan,NNPDF:2024djq,Barontini:2024eii}.
Theory predictions use the pipeline described in Ref.~\cite{Barontini:2023vmr} which is built upon the {\sc\small EKO}~\cite{Candido:2022tld,candido_2025_15655642} evolution code and the {\sc\small PineAPPL} fast grid interface~\cite{Carrazza:2020gss,christopher_schwan_2024_12795745}.
These results always include MHOUs on the theory prediction as discussed in Ref.~\cite{NNPDF:2024dpb},
and, in each case, with NNLO or aN$^3$LO QCD theory, and with or without QED corrections.
QED corrections are included according to Ref.~\cite{NNPDF:2024djq}, and aN$^3$LO QCD corrections following Ref.~\cite{NNPDF:2024nan}, updated with the most recent implementation of heavy quark matching of Ref.~\cite{Ablinger:2024xtt}.
The value of $\alpha_s$ is extracted using the
CRM and the TCM, respectively discussed in Sects.~\ref{subsec:CRM}
and \ref{subsec:TCM}, with the settings discussed in
Sect.~\ref{subsec:comp}, and the same number of replicas used in each closure
test run (see Sect.~\ref{subsec:settings}), namely for the CRM $N_{\text{rep}} =
250$ replicas for each of the 12 values of $\alpha_s$ under consideration, and
$N_{\text{rep}} = 550$ replicas for the TCM. As with the closure test, these
are the numbers of replicas before the post-fit selection used to filter
outliers~\cite{Ball:2021leu}. The number of replicas are chosen
such that the finite-size uncertainty as estimated through bootstrapping is less
than one permille on the central value of the extracted $\alpha_s$.
Uncertainties are determined both
as one-sigma and 68\% CL intervals from the replica sample, and for the TCM also
using the analytic formula Eq.~(\ref{eq:TCM_final_2}), with results always
agreeing to within the number of decimal figures shown in the table (i.e.\ at the
permille level). Because we include both experimental uncertainties and
MHOUs, and we simultaneously determine $\alpha_s$ and the PDFs, the resultant uncertainty
includes methodological, experimental and theoretical (nuclear and MHO)
uncertainties,
though not the systematic uncertainty related to the positivity  bias, detected in the closure test of
Sect.~\ref{subsec:ctresults} and further discussed below.
Results are collected in Table~\ref{tab:alphas_baseline}.

\paragraph{Methodology variations.}
Inspection of  Table~\ref{tab:alphas_baseline} shows that the TCM and CRM
results are always in agreement, with differences in central values
and uncertainties at the permille level.
We have also repeated all determinations using the deprecated
method
discussed at the end of Sect.~\ref{subsec:CRM} and based on
Eqs.~(\ref{eq:tzero})--(\ref{eq:assimple}), neglecting the correlation
between  $\alpha_s$ and the PDF as in
Refs.~\cite{Lionetti:2011pw,Ball:2011us}.
We have verified that this leads to the same central values, also at the permille level, but to an underestimate of the uncertainty by
up to 40\%.
We have also recomputed all values while excluding the
MHOUs~\cite{NNPDF:2024dpb}, as is done at NNLO by all other
groups. This leads to an underestimate of the uncertainty which
at NNLO can be
up to a factor of two, with an associated shift in the central value of about
one sigma: upwards at NNLO and downwards at aN$^3$LO. Specifically we
find that the uncertainty on the pure QCD result increases from
$\pm0.0004$ to $\pm0.0008$ at NNLO and from $\pm0.0006$ to
$\pm0.0007$ at aN$^3$LO, suggesting that the MHOU is about $\pm0.0007$
at NNLO and $\pm0.0004$ at aN$^3$LO.

\begin{table}[t]
  \centering
  \renewcommand{\arraystretch}{1.4}
\small
\begin{spacing}{1.3}

\begin{tabularx}{\textwidth}{Xcc}
  \toprule
Perturbative order 
    & TCM  & CRM \\
  \midrule
NNLO$_{\rm QCD}$
&
$0.1195 \pm 0.0008$
&
$0.1203 \pm 0.0008$
\\
NNLO$_{\rm QCD}\otimes$NLO$_{\rm QED}$
&
$0.1200 \pm 0.0008$
&
$0.1204 \pm 0.0010$
\\
aN$^3$LO$_{\rm QCD}$
&
$0.1186 \pm 0.0009$
&
$0.1191 \pm 0.0009$
\\
aN$^3$LO$_{\rm QCD}\otimes$NLO$_{\rm QED}$
&
$0.1187 \pm 0.0009$
&
$0.1194 \pm 0.0008$
\\
\bottomrule
\end{tabularx}
\end{spacing}

    \vspace{0.2cm}
  \caption{Same as Table~\ref{tab:alphas_baseline} but removing the positivity constraint on physical observables.
}
  \label{tab:alphas_baseline_nopos}
\end{table}

\paragraph{Perturbative stability and QED corrections.}
The results shown in Table~\ref{tab:alphas_baseline} show
perturbative stability: the value of $\alpha_s$ decreases as the
perturbative order increases, as previously observed when going from NLO to
NNLO~\cite{Ball:2018iqk}, but  the results at two subsequent orders
agree at the one sigma
level, as they ought to given that MHOUs are included. Indeed, if
MHOUs are not included the NNLO and aN$^3$LO values (CRM, pure QCD) become
respectively $\alpha_s(m_Z)=0.1205\pm0.0004$ and
$\alpha_s(m_Z)=0.1187\pm0.0006$, and hence disagree at  the four--five
sigma level.
Despite the fact that MHOUs contribute substantially to the overall uncertainty,
the total uncertainty does not decrease when going from NNLO to aN$^3$LO.
This is unsurprising as at aN$^3$LO corrections are only included for
perturbative evolution~\cite{Candido:2022tld,candido_2025_15655642} and deep-inelastic coefficient functions~\cite{Candido:2024rkr,candido_2023_8066034}
while for all hadronic
processes, which carry substantial weight in determining $\alpha_s$, partonic
cross-sections are still computed at NNLO with corresponding MHOUs.

The inclusion of QED corrections has the effect of increasing the
value of $\alpha_s$ by a small but non-negligible amount. This can
most likely be understood as a consequence of the fact that the photon
PDF subtracts momentum from the gluon, and the ensuing slight suppression of the
gluon is compensated by a slightly larger value of $\alpha_s$.
It is important to observe that, to the best of our knowledge, the effect of QED corrections is not included in any other simultaneous determination of $\alpha_s$ and PDFs, and, moreover, the associated uncertainty is clearly not included in QCD scale variation and therefore routinely neglected.

\paragraph{Impact of positivity.}
We have found from the closure test analysis of Sect.~\ref{sec:closure_tests} that imposing positivity constraints leads to a bias in
the extracted value of $\alpha_s(m_Z)$.
Therefore, we repeated the determinations shown in Table~\ref{tab:alphas_baseline}, but now removing this positivity constraint.
In this case, the CRM result corresponding to  the outer $\alpha_s$ values
become somewhat unstable: specifically we have verified
that the 68\% CL and one-sigma uncertainties  are
significantly different, and we have traced this to the presence of
outliers in the replica distribution.
We have consequently run two batches according to the method of
Ref.~\cite{Ball:2018iqk}: for each data replica two fits are
performed, and that leading to the best loss is chosen. However,
the non-Gaussianity persists also in this case.

Results for the extraction of $\alpha_s$ when the positivity constraints are not imposed
are shown in  Table~\ref{tab:alphas_baseline_nopos}, with  the same
theory settings as in Table~\ref{tab:alphas_baseline}.
It is clear that just like in the closure test, removing positivity constraints leads to  a downward shift of the $\alpha_s$ value.
As discussed in
Sect.~\ref{subsec:ctresults} the effect of positivity may be
understood as a consequence of the non-Gaussian nature of
uncertainties in the vicinity of kinematic boundaries. However, experimental uncertainties are delivered as Gaussian and consequently
we cannot easily correct for this. Moreover, because the TCM
and the CRM in the absence of positivity no longer agree,
it is not easy to estimate reliably the
the size of the bias. We have therefore conservatively
taken the difference
between the TCM result with and without positivity as an extra source
of uncertainty. We use the TCM result since, in most cases, the shift due to positivity is larger than the corresponding shift of the CRM result. This source of
uncertainty is considered to reflect a non-Gaussian bias, and thus added linearly.
Also, since relaxing positivity in the TCM always produces a downward shift, this contribution is added only to the lower uncertainty, resulting in an asymmetric overall uncertainty.

\begin{figure}[t]
  \centering
  \includegraphics[width=0.70\textwidth]{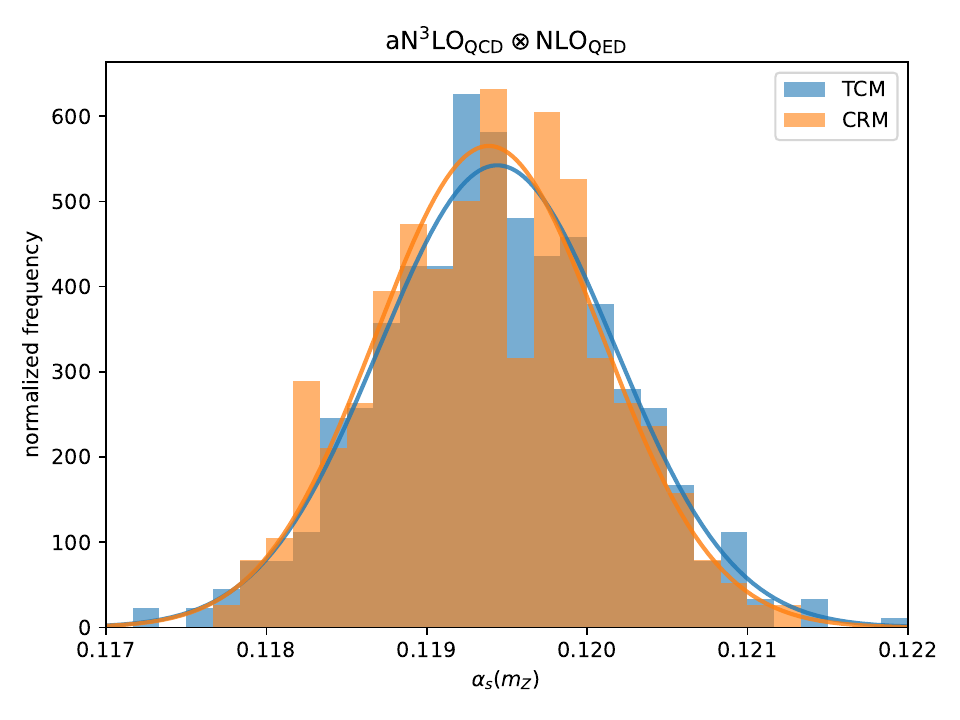}
  \caption{Histogram of the values of the $N_{\rm rep}$ best-fit values
    $\alpha_s^{(k)}$ obtained with the TCM and CRM when applied to the experimental data entering the NNPDF4.0 global fit.
    In both cases, results shown correspond to the fits carried out at aN$^3$LO$_{\rm QCD}\otimes$NLO$_{\rm QED}$
    accuracy and accounting for the positivity of physical observables, see the bottom row of Table~\ref{tab:alphas_baseline}
    for the corresponding central values and 68\% CL uncertainties on $\alpha_s(m_Z)$. The curves are Gaussian fits to the two distributions.
  }
  \label{fig:alphas_histo}
\end{figure}

\paragraph{Final results.}
The histogram of the ensemble of  $\alpha_s$ replica values obtained using the CRM and TCM in the fits with aN$^3$LO$_{\rm QCD}\otimes$NLO$_{\rm QED}$ accuracy and accounting for the positivity constraints is displayed in Fig.~\ref{fig:alphas_histo}.
It is clear that the distributions are both Gaussian and in excellent agreement. We take as our best value for $\alpha_s$ and its
uncertainty that obtained with the TCM, which is based on a larger number
of replicas. The final uncertainty on this value
is determined by adding linearly to
the lower uncertainty the difference between the TCM results with
and without positivity at the corresponding perturbative order. The final results are collected in the last column of
Table~\ref{tab:alphas_baseline}.

\begin{table}[t]
  \centering
  \renewcommand{\arraystretch}{1.4}
\small
\begin{spacing}{1.3}
\begin{tabularx}{\textwidth}{XXcc}
  \toprule
{Perturbative order} & {Theory setting} & TCM  & CRM \\
\midrule
NNLO$_{\rm QCD}$ & expanded solution 
&
$0.1195 \pm 0.0007$
&
$0.1196 \pm 0.0006$
\\

aN$^3$LO$_{\rm QCD}$ & expanded solution
&
$0.1192 \pm 0.0007$
&
$0.1194 \pm 0.0007$
\\
\midrule
NNLO$_{\rm QCD}$ & exp. covmat replicas
&
$0.1199 \pm 0.0007$
&
$0.1199 \pm 0.0006$
\\
NNLO$_{\rm QCD}\otimes$NLO$_{\rm QED}$
 & exp. covmat replicas
&
$0.1202 \pm 0.0006$
&
$0.1201 \pm 0.0006$
\\

aN$^3$LO$_{\rm QCD}$ & exp. covmat replicas
&
$0.1192 \pm 0.0007$
&
$0.1191 \pm 0.0007$\\
aN$^3$LO$_{\rm QCD}\otimes$NLO$_{\rm QED}$
& exp. covmat replicas
&
$0.1194 \pm 0.0007$
&
$0.1195 \pm 0.0007$\\

\bottomrule
\end{tabularx}
\end{spacing}

    \vspace{0.2cm}
  \caption{Results for  the determination of $\alpha_s$ obtained with two variations of
    methodological settings: using the expanded instead of the exact
    solution of evolution equations, and using the experimental
    covariance matrix instead of the $t_0$ covariance matrix for the data
    replica generation (see text).
    For the former, we only compare results of the fits with NNLO$_{\rm QCD}$ and aN$^3$LO$_{\rm QCD}$,
    given that the inclusion of QED corrections requires the use of
    the
    exact solution~\cite{NNPDF:2024djq}.}
  \label{tab:alphas_var}
\end{table}

\begin{figure}[t]
  \centering
  \includegraphics[width=0.75\textwidth]{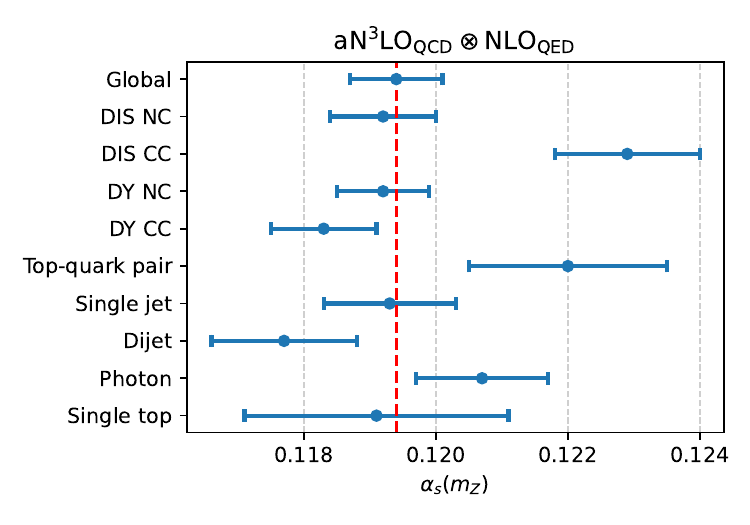}
  \caption{The values of $\alpha_s(m_Z)$ extracted at aN$^3$LO$_{\rm
      QCD}\otimes$NLO$_{\rm QED}$  accuracy from the TCM applied to
  the partial $\chi^2$ evaluated for separate groups of processes.
   In all cases, uncertainties shown correspond to 68\% CL intervals.
  The dashed vertical line corresponds to the best-fit value obtained from the global dataset.}
  \label{fig:alphas_process_sensitivity}
\end{figure}

\subsection{Methodological stability}
\label{sec:variations}

\paragraph{Solution of evolution equations.}
In our default determination of $\alpha_s$, the QCD and
QCD$\otimes$QED
evolution
equations are solved in the same way. This requires using an exact
solution, rather than the expanded solution which was previously used
by us for pure QCD evolution in Refs.~\cite{Ball:2021leu,NNPDF:2024nan},
because construction of the expanded solution for QCD$\otimes$QED
evolution is problematic: see
Ref.~\cite{NNPDF:2024djq} for a detailed discussion. Differences
between the two methods are subleading in the QCD expansion. In
Table~\ref{tab:alphas_var} we show the results obtained by switching to
the expanded solution in the pure QCD determinations at NNLO and
aN$^3$LO. Results are shown to change by less than half a sigma at NNLO,
and to be essentially unchanged at aN$^3$LO. The decrease of the NNLO
result when using truncated evolution reduces the difference between
NNLO and aN$^3$LO by a factor of two.

\begin{table}[t]
  \centering
  \renewcommand{\arraystretch}{1.4}
\small
\begin{spacing}{1.3}
\begin{tabularx}{\textwidth}{Xcccc}
  \toprule
Determination & Perturbative accuracy &   Dataset  &  $\alpha_s(m_Z)$  & Ref.\\
  \midrule
NNPDF4.0 & aN$^3$LO$_{\rm QCD}\otimes$NLO$_{\rm QED}$ & Global  &     $0.1194^{+0.0007}_{-0.0014}$  & This work
\\
\midrule
NNPDF3.1  & NNLO$_{\rm QCD}$ & Global   &   $0.1185 \pm 0.0012$ &  \cite{Ball:2018iqk}
\\
MSHT20  & NNLO$_{\rm QCD}$ & Global   &   $0.1171 \pm 0.0014$ & \cite{Bailey:2020ooq}
\\
MSHT20   & aN$^3$LO$_{\rm QCD}$ & Global   &   $0.1170 \pm 0.0016$ & \cite{Cridge:2024exf}
\\
ABMPtt  &  NNLO$_{\rm QCD}$ & Global (no jets)   &   $0.1150\pm 0.0009 $ & \cite{Alekhin:2024bhs}
\\
\midrule
ATLAS $p_T^Z$ 8 TeV  & ${\rm N}^{3}{\rm LO}\otimes {\rm N}^4{\rm LLa}_{\rm QCD}$   & $d\sigma(Z\to \ell^+\ell^-)/dp_T^Z$  & $0.1183 \pm 0.0009$ &     \cite{ATLAS:2023lhg}
\\
CMS jets 13 TeV   &  NNLO$_{\rm QCD}$  &   $d^2\sigma/dp_T^jdy_j$    & $0.1166 \pm 0.0017$ &     \cite{CMS:2021yzl}
\\
\midrule
ALPHA 25 (lattice QCD)  &  -  & - & $0.11873 \pm 0.00056$ &     \cite{Brida:2025giiy}
\\
\midrule
PDG 2024  &   -         & Average  & $0.1180 \pm 0.0009$  & \cite{ParticleDataGroup:2024cfk}\\
PDG 2024 (no lattice QCD)   &   -         & Average excl. lattice & $0.1175 \pm 0.0010$ & \cite{ParticleDataGroup:2024cfk}\\
\bottomrule
\end{tabularx}
\end{spacing}

    \vspace{0.2cm}
  \caption{Comparison of the results of the NNPDF4.0 determination of $\alpha_s(m_Z)$ presented
  in this work (first row) with other recent determinations of the strong coupling.
  See Fig.~\ref{fig:alphas_nnpdf40_summary} for a graphical representation of the results.
}
  \label{tab:tab_alphas_other}
\end{table}

\begin{figure}[t]
  \centering
  \includegraphics[width=0.70\textwidth]{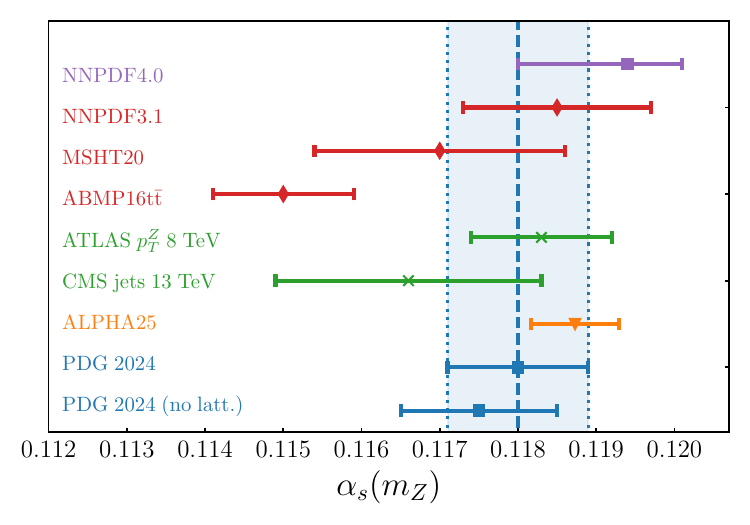}
  \caption{Graphical representation of the results of Table~\ref{tab:tab_alphas_other}.
  For MSHT20, we show the aN$^3$LO QCD result.
  The filled vertical band corresponds to the 2024 PDG average of $\alpha_s(m_Z)=0.1180 \pm 0.0009$.
  }
  \label{fig:alphas_nnpdf40_summary}
\end{figure}

\paragraph{Data replica generation.}

When generating Monte Carlo data replicas (see Sect.~\ref{subsec:CRM})
one may choose to use the experimental covariance matrix, or the $t_0$
covariance matrix, as discussed in Sect.~\ref{subsec:settings}, where
it was mentioned that independence of results on this choice was
explicitly checked. In Table~\ref{tab:alphas_var}  we also show
results obtained by performing this methodological variation.
Comparing to Table~\ref{tab:alphas_baseline} shows that indeed results are all but unaffected by this choice.

\paragraph{Value of the top quark mass.}
Our global dataset includes top production data, for which the
theoretical predictions are sensitive to the value of the top quark mass.
In our default NNPDF4.0 determination~\cite{Ball:2021leu} we adopt the value $m_t=$172.5~GeV for the pole top mass.
We have repeated our NNLO pure QCD determination with $m_t=$175.0~GeV and $m_t=$170.0~GeV. This
corresponds to a variation of almost four times the
PDG pole mass
uncertainty of $\Delta m_t=0.7$~GeV~\cite{ParticleDataGroup:2024cfk}.
Within this wide range we find that the value of $\alpha_s$ changes by $\Delta \alpha=0.0004$ at NNLO and $\Delta\alpha=0.0001$ at aN$^3$LO, increasing with increasing top mass. We conclude that our result is essentially independent of the value of the top quark mass.
This is likely a consequence of the fact that the top pair production data constitute a relatively small subset of our global dataset, in particular when
compared to other gluon-sensitive measurements such as single-inclusive jets and dijets.

\subsection{Comparison to other determinations}
\label{sec:dataset}

\paragraph{Dataset dependence.}
Fig.~\ref{fig:alphas_process_sensitivity} displays the values of $\alpha_s(m_Z)$ extracted at aN$^3$LO$_{\rm QCD}\otimes$NLO$_{\rm
  QED}$  accuracy using the TCM  applied to the partial $\chi^2$ evaluated for separate (exclusive) groups of processes.
In all cases, uncertainties shown correspond to 68\% CL intervals.
The values shown  give an indication of the $\alpha_s$ preferred by
individual processes. However, they cannot be understood as the best-fit values
associate to that process~\cite{Forte:2020pyp}, and  in particular the
global $\alpha_s$ value does not correspond to their weighted mean.
This is not only because these values neglect
correlations between different processes, but also because
for each process there exist in general values of $\alpha_s$ that
give a better fit to it while giving the same quality fit to the rest
of the dataset~\cite{Ball:2018iqk,Forte:2020pyp}.
This said, the qualitative indication coming from
Fig.~\ref{fig:alphas_process_sensitivity} is that charged-current
deep--inelastic structure functions, direct photon production, and top quark pair production data prefer a larger value
of $\alpha_s(m_Z)$, while Drell-Yan  charged current and dijet cross-sections instead prefer a lower one.

We have furthermore verified that using the NNPDF3.1-like dataset (as
defined in Ref.~\cite{Ball:2021leu}) and the same theory settings as
in Ref.~\cite{Ball:2018iqk}, namely pure QCD NNLO theory without MHOUs
and with expanded
solution of the evolution equation, the value of $\alpha_s$ extracted
with the TCM is $\alpha_s(m_Z)=0.1189\pm0.0005$. This is to be compared to
the value $\alpha_s(m_Z)=0.1185\pm0.0005$ obtained with the CRM in that
reference.
Using instead the exact solution and including MHOUs the NNLO
value from the NNPDF3.1-like dataset is
$\alpha_s(m_Z)=0.1188\pm0.0006$.
This implies that a substantial part of the
difference between the value of $\alpha_s$ of
Ref.~\cite{Ball:2018iqk}, and the rather higher NNLO value of
Tab.~\ref{tab:alphas_baseline} is due to the much larger weight of LHC
data in the NNPDF4.0 dataset, and not to any methodological differences, and in particular not at all to differences between the
NNPDF3.1 and NNPDF4.0 methodology. The determination of $\alpha_s$ from
Ref.~\cite{Ball:2018iqk} was assigned an extra MHOU uncertainty of
$\pm0.0011$, estimated as half the shift between the NLO and NNLO
$\alpha_s$ values, as the formalism of Refs.~\cite{NNPDF:2019vjt,NNPDF:2019ubu}
for the inclusion of MHOUs was not yet available at the
time. Interestingly, the MHOU on the NNLO result  determined here,
$\pm0.0007$, see Sec.~\ref{sec:baseline}, is
smaller by almost a factor of two.

\paragraph{Other $\alpha_s$ determinations}
Table~\ref{tab:tab_alphas_other} displays the comparison of the results of the NNPDF4.0 extraction of $\alpha_s(m_Z)$ presented
in this work, based on aN$^3$LO$_{\rm QCD}\otimes$NLO$_{\rm QED}$  theory calculations
and accounting for MHOUs, with other recent determinations of the strong coupling
jointly with PDFs.
Specifically, we compare with the MSHT20 NNLO and aN$^3$LO determinations, the ABMPtt updated analysis including
differential top quark data,
as well as with our previous NNLO determination based on  NNPDF3.1.
We also include the two single most precise determinations
performed by ATLAS and CMS and based on the $p_T$ distributions of $Z$ bosons at 8 TeV
and on the double-differential single-inclusive jet cross-sections at
13 TeV respectively, though only in the latter PDFs are determined simultaneously
with the strong coupling.\footnote{A recent extension of the CMS analysis of~\cite{CMS:2021yzl} combines their inclusive jet production data at 2.76, 7, 8, and 13 TeV with inclusive HERA structure functions to yield $\alpha_s(m_Z)=0.1176_{-0.0016}^{+0.0014}$~\cite{CMS:2024rkg}.}

We finally display the recent lattice result~\cite{Brida:2025giiy}
from the ALPHA collaboration, which is the single most precise
determination, and the latest published  PDG averages, both
global and not including lattice QCD input.
See Fig.~\ref{fig:alphas_nnpdf40_summary} for the corresponding graphical representation of the results, where for MSHT20 we only display the aN$^3$LO QCD result.

All results shown in Table~\ref{tab:tab_alphas_other} and
Fig.~\ref{fig:alphas_nnpdf40_summary} overlap within uncertainties among themselves,
except the ABMPtt value. Note, however, that the latter presents
a simultaneous determination of $\alpha_s$ and the $\overline{\rm MS}$
top mass; if the
PDG value of the top mass is used, then a higher value of
$\alpha_s$ consistent with the PDF average is obtained~\cite{Alekhin:2024bhs}.

\section{Summary and outlook}
\label{sec:summary}

We have presented an extraction of $\alpha_s(m_Z)$ with high precision
and accuracy: the width of the (asymmetric) uncertainty band in our
determination is the same as that of the PDG combination that
excludes lattice data.
Our determination of $\alpha_s$ has several unique features, all of
which are implemented to the best of our  knowledge for the first time
in a simultaneous determination of $\alpha_s$ and PDFs:
\begin{itemize}
  \item The extraction is performed using both frequentist Monte Carlo
    resampling, and Bayesian inference.
  \item Its methodology is validated by a closure test.
  \item Uncertainties include the MHOUs on the processes used for PDF
    determination both at NNLO and aN$^3$LO.
  \item Effects of mixed QCD$\otimes$QED evolution and the photon PDF
    are accounted for.
\end{itemize}

All of these are important for the reliability of the
results. Specifically, without the closure test analysis it would have
been impossible to detect the bias due to imposing positivity constraints. Without MHOUs the uncertainty on $\alpha_s$
would have been underestimated by up to a factor two. The inclusion of
QED corrections affects the central value of $\alpha_s$ at the level of a few permille. We have no reason to believe that these effects would not have
a comparable impact if they were studied or included in other
simultaneous determinations of PDFs and $\alpha_s$.

It will be interesting in the future to use the methods deployed in
this work to carry out joint determinations of PDFs with
other physical parameters in addition to $\alpha_s(m_Z)$, such as the
top quark mass, and to validate them with closure tests.
Also, with the availability of more data it might be interesting to carry out  $\alpha_s(Q)$ extractions in separate bins of $Q$, in order to constrain new physics scenarios which may distort the scale dependence  of $\alpha_s(Q)$ in comparison to the standard model prediction~\cite{CMS:2014mna}.

\begin{center}
\rule{5cm}{.1pt}
\end{center}
\bigskip
The NNPDF4.0 PDF sets used for this work are
available, in the {\sc\small LHAPDF} format~\cite{Buckley:2014ana}, through the NNPDF website:
\begin{center}
\url{https://nnpdf.mi.infn.it/nnpdf4-0-alphas/}
\end{center}
Specifically, we release NNLO and aN$^3$LO QCD sets, without and with QED corrections, for all values of $\alpha_s(m_Z)$ used
for the present determination. All
sets are composed of $N_{\rm rep}=200$ replicas. In all cases MHOUs
are included, and multiplicative correlated uncertainties are
determined using a fixed $t_0$ matrix corresponding to the
PDF set at the best $\alpha_s$ which is indicated in the set name,
for the reason explained in Sect.~\ref{subsec:settings}. The replicas
are correlated, meaning that replicas with the same index corresponding to
different values of $\alpha_s$ are all fitted to the same underlying
data replica, see Sect.~\ref{subsec:CRM}.

They are denoted as follows:

\begin{itemize}

\item NNLO QCD (+ QED effects)

{\tt NNPDF40\_nnlo\_as\_01140\_mhou\_t0120}$\quad$ ({\tt NNPDF40\_nnlo\_as\_01140\_qed\_mhou\_t0120})\\
{\tt NNPDF40\_nnlo\_as\_01150\_mhou\_t0120}$\quad$ ({\tt NNPDF40\_nnlo\_as\_01150\_qed\_mhou\_t0120})\\
{\tt NNPDF40\_nnlo\_as\_01160\_mhou\_t0120}$\quad$ ({\tt NNPDF40\_nnlo\_as\_01160\_qed\_mhou\_t0120})\\
{\tt NNPDF40\_nnlo\_as\_01170\_mhou\_t0120}$\quad$ ({\tt NNPDF40\_nnlo\_as\_01170\_qed\_mhou\_t0120})\\
{\tt NNPDF40\_nnlo\_as\_01180\_mhou\_t0120}$\quad$ ({\tt NNPDF40\_nnlo\_as\_01180\_qed\_mhou\_t0120})\\
{\tt NNPDF40\_nnlo\_as\_01190\_mhou\_t0120}$\quad$ ({\tt NNPDF40\_nnlo\_as\_01190\_qed\_mhou\_t0120})\\
{\tt NNPDF40\_nnlo\_as\_01200\_mhou\_t0120}$\quad$ ({\tt NNPDF40\_nnlo\_as\_01200\_qed\_mhou\_t0120})\\
{\tt NNPDF40\_nnlo\_as\_01210\_mhou\_t0120}$\quad$ ({\tt NNPDF40\_nnlo\_as\_01210\_qed\_mhou\_t0120})\\
{\tt NNPDF40\_nnlo\_as\_01220\_mhou\_t0120}$\quad$ ({\tt NNPDF40\_nnlo\_as\_01220\_qed\_mhou\_t0120})\\
{\tt NNPDF40\_nnlo\_as\_01230\_mhou\_t0120}$\quad$ ({\tt NNPDF40\_nnlo\_as\_01230\_qed\_mhou\_t0120})\\
{\tt NNPDF40\_nnlo\_as\_01240\_mhou\_t0120}$\quad$ ({\tt NNPDF40\_nnlo\_as\_01240\_qed\_mhou\_t0120})\\
{\tt NNPDF40\_nnlo\_as\_01250\_mhou\_t0120}$\quad$ ({\tt NNPDF40\_nnlo\_as\_01250\_qed\_mhou\_t0120})\\

\item aN$^3$LO QCD (+ QED effects)

{\tt NNPDF40\_an3lo\_as\_01140\_mhou\_t0119}$\quad$ ({\tt NNPDF40\_an3lo\_as\_01140\_qed\_mhou\_t0119})\\
{\tt NNPDF40\_an3lo\_as\_01150\_mhou\_t0119}$\quad$ ({\tt NNPDF40\_an3lo\_as\_01150\_qed\_mhou\_t0119})\\
{\tt NNPDF40\_an3lo\_as\_01160\_mhou\_t0119}$\quad$ ({\tt NNPDF40\_an3lo\_as\_01160\_qed\_mhou\_t0119})\\
{\tt NNPDF40\_an3lo\_as\_01170\_mhou\_t0119}$\quad$ ({\tt NNPDF40\_an3lo\_as\_01170\_qed\_mhou\_t0119})\\
{\tt NNPDF40\_an3lo\_as\_01180\_mhou\_t0119}$\quad$ ({\tt NNPDF40\_an3lo\_as\_01180\_qed\_mhou\_t0119})\\
{\tt NNPDF40\_an3lo\_as\_01190\_mhou\_t0119}$\quad$ ({\tt NNPDF40\_an3lo\_as\_01190\_qed\_mhou\_t0119})\\
{\tt NNPDF40\_an3lo\_as\_01200\_mhou\_t0119}$\quad$ ({\tt NNPDF40\_an3lo\_as\_01200\_qed\_mhou\_t0119})\\
{\tt NNPDF40\_an3lo\_as\_01210\_mhou\_t0119}$\quad$ ({\tt NNPDF40\_an3lo\_as\_01210\_qed\_mhou\_t0119})\\
{\tt NNPDF40\_an3lo\_as\_01220\_mhou\_t0119}$\quad$ ({\tt NNPDF40\_an3lo\_as\_01220\_qed\_mhou\_t0119})\\
{\tt NNPDF40\_an3lo\_as\_01230\_mhou\_t0119}$\quad$ ({\tt NNPDF40\_an3lo\_as\_01230\_qed\_mhou\_t0119})\\
{\tt NNPDF40\_an3lo\_as\_01240\_mhou\_t0119}$\quad$ ({\tt NNPDF40\_an3lo\_as\_01240\_qed\_mhou\_t0119})\\
{\tt NNPDF40\_an3lo\_as\_01250\_mhou\_t0119}$\quad$ ({\tt NNPDF40\_an3lo\_as\_01250\_qed\_mhou\_t0119})\\

\end{itemize}

In addition, we also release sets in which multiplicative correlated
uncertainties are determined in each case using the $t_0$ matrix corresponding
to the respective value of $\alpha_s$. Unlike the above sets, for these sets the
replicas are not correlated across different values of
$\alpha_s$. These should not be used for $\alpha_s$ determination, for
the reasons  discussed in
Sect.~\ref{subsec:settings}. However, if $\alpha_s$ is fixed as an external
parameter they provide the most accurate prediction. They are denoted as

\begin{itemize}

\item NNLO QCD (+ QED effects)

{\tt NNPDF40\_nnlo\_mhou\_as\_01140}$\quad$ ({\tt NNPDF40\_nnlo\_mhou\_as\_01140\_qed})\\
{\tt NNPDF40\_nnlo\_mhou\_as\_01150}$\quad$ ({\tt NNPDF40\_nnlo\_mhou\_as\_01150\_qed})\\
{\tt NNPDF40\_nnlo\_mhou\_as\_01160}$\quad$ ({\tt NNPDF40\_nnlo\_mhou\_as\_01160\_qed})\\
{\tt NNPDF40\_nnlo\_mhou\_as\_01170}$\quad$ ({\tt NNPDF40\_nnlo\_mhou\_as\_01170\_qed})\\
{\tt NNPDF40\_nnlo\_mhou\_as\_01180}$\quad$ ({\tt NNPDF40\_nnlo\_mhou\_as\_01180\_qed})\\
{\tt NNPDF40\_nnlo\_mhou\_as\_01190}$\quad$ ({\tt NNPDF40\_nnlo\_mhou\_as\_01190\_qed})\\
{\tt NNPDF40\_nnlo\_mhou\_as\_01200}$\quad$ ({\tt NNPDF40\_nnlo\_mhou\_as\_01200\_qed})\\
{\tt NNPDF40\_nnlo\_mhou\_as\_01210}$\quad$ ({\tt NNPDF40\_nnlo\_mhou\_as\_01210\_qed})\\
{\tt NNPDF40\_nnlo\_mhou\_as\_01220}$\quad$ ({\tt NNPDF40\_nnlo\_mhou\_as\_01220\_qed})\\
{\tt NNPDF40\_nnlo\_mhou\_as\_01230}$\quad$ ({\tt NNPDF40\_nnlo\_mhou\_as\_01230\_qed})\\
{\tt NNPDF40\_nnlo\_mhou\_as\_01240}$\quad$ ({\tt NNPDF40\_nnlo\_mhou\_as\_01240\_qed})\\
{\tt NNPDF40\_nnlo\_mhou\_as\_01250}$\quad$ ({\tt NNPDF40\_nnlo\_mhou\_as\_01250\_qed})\\

\item aN$^3$LO QCD (+ QED effects)

{\tt NNPDF40\_an3lo\_mhou\_as\_01140}$\quad$ ({\tt NNPDF40\_an3lo\_mhou\_as\_01140\_qed})\\
{\tt NNPDF40\_an3lo\_mhou\_as\_01150}$\quad$ ({\tt NNPDF40\_an3lo\_mhou\_as\_01150\_qed})\\
{\tt NNPDF40\_an3lo\_mhou\_as\_01160}$\quad$ ({\tt NNPDF40\_an3lo\_mhou\_as\_01160\_qed})\\
{\tt NNPDF40\_an3lo\_mhou\_as\_01170}$\quad$ ({\tt NNPDF40\_an3lo\_mhou\_as\_01170\_qed})\\
{\tt NNPDF40\_an3lo\_mhou\_as\_01180}$\quad$ ({\tt NNPDF40\_an3lo\_mhou\_as\_01180\_qed})\\
{\tt NNPDF40\_an3lo\_mhou\_as\_01190}$\quad$ ({\tt NNPDF40\_an3lo\_mhou\_as\_01190\_qed})\\
{\tt NNPDF40\_an3lo\_mhou\_as\_01200}$\quad$ ({\tt NNPDF40\_an3lo\_mhou\_as\_01200\_qed})\\
{\tt NNPDF40\_an3lo\_mhou\_as\_01210}$\quad$ ({\tt NNPDF40\_an3lo\_mhou\_as\_01210\_qed})\\
{\tt NNPDF40\_an3lo\_mhou\_as\_01220}$\quad$ ({\tt NNPDF40\_an3lo\_mhou\_as\_01220\_qed})\\
{\tt NNPDF40\_an3lo\_mhou\_as\_01230}$\quad$ ({\tt NNPDF40\_an3lo\_mhou\_as\_01230\_qed})\\
{\tt NNPDF40\_an3lo\_mhou\_as\_01240}$\quad$ ({\tt NNPDF40\_an3lo\_mhou\_as\_01240\_qed})\\
{\tt NNPDF40\_an3lo\_mhou\_as\_01250}$\quad$ ({\tt NNPDF40\_an3lo\_mhou\_as\_01250\_qed})\\

\end{itemize}

\bigskip
\paragraph{Acknowledgements:}
We thank all the members of the NNPDF collaboration for numerous
discussions and support during the course of this work.
We are grateful to Thomas Cridge, David D'Enterria, Maria-Vittoria Garzelli,
Lucian Harland-Lang, Sven Moch, Klaus Rabbertz, and Robert Thorne for
discussions.

R.D.B and R.S. thank the Science and Technology Facilities Council
(STFC) for support via grant awards ST/T000600/1 and ST/X000494/1.
S.F. is partly funded by the European Union
NextGeneration EU program, NRP Mission 4 Component 2 Investment 1.1–
MUR PRIN 2022CUPG53D23001100006 through the Italian Ministry of
University and Research.
F.H. has been supported by the Academy of Finland project 358090 and was funded as a part of the Center of Excellence in Quark Matter of the Academy
of Finland, project 346326.
E.R.N. is supported by the Italian Ministry of University and Research (MUR)
through the “Rita Levi-Montalcini” Program.
J.R. is partially supported by NWO, the Dutch Research Council.

\bibliography{nnpdf40alphas}

\end{document}